\PassOptionsToPackage{pdfpagelabels=false}{hyperref} 
\documentclass[a4paper,fleqn,usenatbib,useAMS]{mnras} 

\usepackage{graphicx}	
\usepackage{amsmath}	
\usepackage{amssymb}	
\usepackage{multicol}   
\usepackage{bm}			
\usepackage{pdflscape}	
\usepackage{subfig}     
\usepackage[toc,page]{appendix}
\usepackage{hhline}
\usepackage{longtable}
\usepackage{supertabular}
\usepackage{silence} 
\usepackage{makecell}
\usepackage{relsize}
\usepackage[normalem]{ulem}
\usepackage{needspace}

\usepackage{todonotes}	

\newcommand{\mcite}[1]{\mbox{\cite{#1}}}
\newcommand{\mcitealt}[1]{\mbox{\citealt{#1}}}
\newcommand{\mcitep}[1]{\mbox{\citep{#1}}}

\newcommand{\kunit}{\text{h}\text{Mpc}^{-1}}
\newcommand{\deltatb}{\delta T_{\mathrm{b}}}
\newcommand{\Deltatb}{\Delta T_{\mathrm{b}}}

\newcommand{\cmfast}{\textit{21cmFAST}}
\newcommand{\simfast}{\textit{SimFast21}}
\newcommand{\cray}{C\textsuperscript{2}-RAY}

\newcommand{\Mmin}{M_{\mathrm{min}}}
\newcommand{\Tvir}{T_{\mathrm{vir}}}
\newcommand{\Rmax}{R_{\mathrm{max}}}
\newcommand{\Zion}{\zeta_{\mathrm{ion}}}
\newcommand{\fcoll}{f_{\mathrm{coll}}}

\newcommand{\eqnref}[1]{Equation~(\ref{#1})}
\newcommand{\secref}{Section~\ref}
\newcommand{\figref}{Figure~\ref}
\newcommand{\tabref}{Table~\ref}

\newcommand{\changes}{}

\usepackage[T1]{fontenc}
\usepackage{ae,aecompl}

\usepackage{newtxtext,newtxmath}

\title[Evaluating 21cm machine learning techniques]{Evaluating machine learning techniques for predicting power spectra from reionization simulations}

\author[W. D. Jennings et al.]{ W. D. Jennings $^{1}$, C. A. Watkinson $^{2}$, F. B. Abdalla $^{1}$, J. D. McEwen $^{3}$
\\
$^{1}$ Department of Physics \& Astronomy, University College London, Gower Street, London WC1E 6BT, UK
\\
$^{2}$ Blackett Laboratory, Imperial College, London, SW7 2AZ, UK
\\
$^{3}$ Mullard Space Science Laboratory, University College London, Surrey RH5 6NT, UK
}

\date{Accepted 2018 November 20. Received 2018 November 20; in original form 2018 September 14}

\pubyear{2018}

\begin{document}
\label{firstpage}
\pagerange{\pageref{firstpage}--\pageref{lastpage}}
\maketitle

\begin{abstract}

Upcoming experiments such as the SKA will provide huge quantities of data. Fast modelling of the high-redshift 21cm signal will be crucial for efficiently comparing these data sets with theory. The most detailed theoretical predictions currently come from numerical simulations and from faster but less accurate semi-numerical simulations. Recently, machine learning techniques have been proposed to emulate the behaviour of these semi-numerical simulations with drastically reduced time and computing cost. We compare the viability of five such machine learning techniques for emulating the 21cm power spectrum of the publicly-available code \simfast{}. Our best emulator is a multilayer perceptron with three hidden layers, reproducing \simfast{} power spectra $10^8$ times faster than the simulation with 4\% mean squared error averaged across all redshifts and input parameters. The other techniques (interpolation, Gaussian processes regression, and support vector machine) have slower prediction times and worse prediction accuracy than the multilayer perceptron. All our emulators can make predictions at any redshift and scale, which gives more flexible predictions but results in significantly worse prediction accuracy at lower redshifts. We then present a proof-of-concept technique for mapping between two different simulations, exploiting our best emulator's fast prediction speed. We demonstrate this technique to find a mapping between \simfast{} and another publicly-available code \cmfast{}. We observe a noticeable offset between the simulations for some regions of the input space. Such techniques could potentially be used as a bridge between fast semi-numerical simulations and accurate numerical radiative transfer simulations.

\end{abstract}



\section{Introduction}
The Dark Ages of the Universe ended when the first stars and galaxies begin to form. Radiation from these sources ionized the surrounding matter, eventually giving rise to bubbles of ionized hydrogen. The size, shape and clustering properties of these bubbles contain valuable information about how our Universe evolved during these otherwise obscure times. Direct observation of these bubbles requires us to distinguish between ionized regions and neutral regions. The most promising probe for this is the 21cm hyperfine transition of hydrogen, which is emitted exclusively by neutral hydrogen during the proton-electron interaction. Measurements of the 21cm signal on the sky give us an image of the neutral hydrogen in the Universe and, by tracing this signal signal back through time, we can extend these images into three-dimensional maps.

Observational difficulties have so far prevented us from creating such three-dimensional maps. The signal is much weaker than other foreground sources at similar frequencies and it is difficult to extract the actual 21cm signal from these foregrounds. Past and ongoing experiments such as Murchison Widefield Array\footnote{http://www.mwatelescope.org/telescope} (MWA, \mcitealt{mwa}), the Low Frequency
Array\footnote{http://www.lofar.org/} (LOFAR, \mcitealt{LOFAR}), and the Precision Array for Probing the Epoch of Reionization\footnote{http://eor.berkeley.edu/} (PAPER, \mcitealt{PAPER}) have begun to place limits on the overall intensity of the signal. Upcoming experiments such as the Hydrogen Epoch of Reionization Array\footnote{http://reionization.org/} (HERA, \mcitealt{HERA}) and the Square Kilometre Array\footnote{https://www.skatelescope.org/} (SKA, \mcitealt{ska}) will be able to provide more detailed measurements and should allow us to make first parameter constraints for our models.

Theoretical modelling of the 21cm signal involves answering questions about the reionization processes: what were the main sources of ionizing photons; when did reionization start; how long did it last? The most detailed theoretical predictions are currently from numerical and semi-numerical simulations. The most likely reionization scenarios can be extracted by comparing such simulations to data, most efficiently by combining fast approximate semi-numerical simulations with sampling methods such as MCMC to perform parameter estimation (\mcitealt{21CMMC2015}, \changes{\mcitealt{Hassan2017}, \mcitealt{Liu2016}, \mcitealt{Pober2016}, \mcitealt{Greig2016}, \mcitealt{Greig2018}}). Two semi-numerical simulations are \simfast{} \mcitep{Simfast21} and \cmfast{} \mcitep{21cmFast}, which generate three-dimensional realisations of the 21cm signal. Much work has gone into finding efficient summary statistics for the simulation outputs. Common summary statistics are the power spectrum and its higher-order equivalent the bispectrum (\mcitealt{Shimabukuro2016Bispectrum}, \mcitealt{Watkinson2017}, \mcitealt{suman2018} and \mcitealt{Watkinson2018}). Both statistics contain information about the clustering properties of ionized hydrogen bubbles.

Semi-numerical simulations take minutes to hours to run. Recently machine learning techniques have been suggested for a number of uses: to emulate power spectrum outputs quickly from \cmfast{} (\mcitealt{Schmit2018}, \mcitealt{Kern2017}), to derive reionization parameters directly from the 21cm power spectrum (\mcitealt{Shimabukuro2017}), and to derive reionization parameters from 21cm images (\mcitealt{Gillet2018}). In the first application, models are trained to mimic the outputs that would have resulted from an actual simulation. Training involves running a representative sample of actual simulations and learning to mimic their behaviour. After training, the models can make fast power spectrum predictions at any new input points. The ultimate aim of such an approach would be to train models to mimic the more accurate numerical simulations, allowing for more accurate parameter estimation.

In this paper, we evaluate the viability of five machine learning techniques for emulating the 21cm power spectrum from \simfast{}. We analyse the prediction speeds of the resulting emulators and their accuracy across the standard reionization input parameter space. The emulators in \mcitealt{Schmit2018} and \mcitealt{Kern2017} were trained at fixed scales and fixed redshifts. Such emulators make predictions only at these fixed scales and redshifts, so that if other scales or redshifts are desired one must interpolate further. We use the scales and redshifts directly as extra inputs to the trained models, so that they learn to make predictions for any requested scale and redshift. This method is theoretically more flexible but gives rise to poorer prediction accuracy at lower redshifts.

We then use our best emulator candidate to present a proof-of-concept technique for determining a relationship between two different simulations. We demonstrate the technique by finding a mapping between the inputs of \simfast{} and those of \cmfast{} by measuring which inputs result in the most similar output power spectra. This method could potentially be used to bridge between fast semi-numerical simulations and more accurate three-dimensional radiative transfer codes, see for example \cray{} \mcitep{c2ray} and LICORICE \mcitep{Semelin2017, Girish2016}. In our conclusions we comment on the feasibility of using our method for this purpose, both in light of our results and in the context of the known discrepancy between numerical and semi-numerical codes (see for example \mcitealt{Suman2014}).

The rest of the paper is split in to the following sections. In \secref{sec:ReionizationModel} we describe the reionization models used in the simulations. \secref{sec:ML} contains descriptions of the machine learning techniques we used. In \secref{sec:training} we briefly describe the specifics of how our emulators were trained. We present the results of training our emulators in \secref{sec:training results}. \secref{sec:discussion-emulators} is a discussion of the accuracy and speed performance of the different machine learning techniques, and how their performance depends on the input parameters. \secref{sec:bias} contains the proof-of-concept method for mapping between \simfast{} and modified \cmfast{}, using our best emulator.  We end the paper in \secref{sec:conclusions} with our conclusions. For Cosmological parameters, we use $\Omega_{\mathrm{M}} = 0.270$,  $\Omega_{\mathrm{b}} = 0.046$,  $\Omega_{\mathrm{\Lambda}} = 0.730$,  $H_0 = 71.0 \text{km s}^{-1}\text{Mpc}^{-1}$, $n_{\mathrm{s}} = 0.960$, $\sigma_8 = 0.810$, the default parameters in the \simfast{} package\footnote{https://github.com/mariogrs/Simfast21}.

\section{Models of reionization}
\label{sec:ReionizationModel}

The 21cm differential brightness temperature $\deltatb$ is defined as the difference between the measured 21cm brightness temperature and the uniform background CMB brightness temperature. By removing the background CMB temperature, the value of $\deltatb (\vec{r})$ then specifies the extent of 21cm emission ($\deltatb > 0$) or absorption ($\deltatb < 0$). The actual observable for radio interferometers is $\deltatb - \langle \deltatb \rangle$, where $\langle \deltatb \rangle$ is the global reionization signal averaged across the whole sky. \mcite{Furlanetto2006} gives an approximate relationship for the 21cm brightness temperature $\deltatb(\vec{r})$ as

\begin{align}
\label{eqn:deltaTb approx}
\deltatb(\vec{r}) = 27 
& x_{\mathrm{HI}}(\vec{r})\ 
\big[ 1 + \delta(\vec{r}) \big]
\left(\frac{\Omega_{\mathrm{b}} h^2}{0.023}\right)
\left(\frac{0.15}{\Omega_{\mathrm{M}} h^2}\right)^{1/2} \\ 
& \left(1 - \frac{T_{\mathrm{\Gamma}}}{T_{\mathrm{S}}}\right)
\left(\frac{1+z}{10}\right)^{1/2}
\notag \left(\frac{H(z)}{H(z) + \delta_r v_r (\vec{r})}\right)
\ \text{mK}\,.
\end{align}

\noindent This approximation includes the effects of neutral hydrogen fraction $x_{\mathrm{HI}}(\vec{r})$; total matter density contrast $\delta(\vec{r})$; cosmological parameters for the densities of baryonic matter $\Omega_{\mathrm{b}}$ and total matter $\Omega_{\mathrm{M}}$; the CMB temperature $T_{\mathrm{\Gamma}}$; the spin temperature $T_{\mathrm{S}}$ which quantifies the relative populations of neutral hydrogen atoms in the higher and lower energy states; the Hubble parameter $H(z)$; and $\delta_r v_r (\vec{r})$, the radial velocity gradient.

\subsection{Power spectrum for \texorpdfstring{$\deltatb$}{deltatb}}
We train our emulators to reproduce correlations in fluctuations of the differential brightness temperature. Fluctuations in $\deltatb(\vec{r})$ are given by

\begin{equation}
\label{eqn:DeltaT_b def}
\Deltatb(\vec{r}) = \frac{\deltatb(\vec{r}) - \left\langle \deltatb(\vec{r}) \right\rangle}{\left\langle \deltatb(\vec{r}) \right\rangle }\,,
\end{equation}

\noindent where $\langle \deltatb (\vec{r}) \rangle$ is again the global reionization signal measured across the whole sky. The correlation in these fluctuations is the power spectrum

\begin{align}
\label{eqn:power_spectrum_definition}
P_{\Deltatb}(\vec{k})\ \delta_D^3(\vec{k}-\vec{k}') = \frac{1}{\left( 2 \pi \right)^3} \left\langle \widetilde{\Delta} T_{\mathrm{b}}(\vec{k})\ \widetilde{\Delta} T_{\mathrm{b}}^{*}(\vec{k}') \right\rangle\,.
\end{align}

\noindent Here, $\widetilde{\Delta} T_{\mathrm{b}}(\vec{k})$ is the Fourier transform of $\Deltatb(\vec{r})$, and the angular brackets denote an ensemble average.

\subsection{SimFast21}
\label{sec:simfast21}
To generate our three-dimensional 21cm maps, we use the publicly available semi-numerical code \simfast{}\footnote{https://github.com/mariogrs/Simfast21} (version 1.0). We briefly describe the algorithm here. The simulation begins by seeding an initial linear density field onto a three-dimensional grid at very high redshift. This linear density field is evolved using first-order perturbation theory (see \mcitealt{Zeldovich1970}) giving a non-linear density field $\delta(\vec{r})$.

The simulation then finds the highest density regions where the matter will collapse to form luminous structures and thus contribute ionizing photons towards the reionization process. The extent of collapse is calculated from the non-linear density field in two different ways. For the collapse of the largest and most massive regions, \simfast{} explicitly resolves individual dark matter halos using an excursion-set formalism \mcitep{Furlanetto2004b}. This method is only used for the collapse of regions larger than a single pixel which means that halos can be resolved down to $5 \times 10^9 M_{\odot}$ in our simulations. For smaller unresolved regions, \simfast{} uses the approximate ellipsoidal collapse method from \mcite{Sheth2001}: if the mean enclosed density in a region exceeds a theoretical critical value then the region is assumed to collapse. The collapse fraction $\fcoll(\vec{r}, R)$ on decreasing scales $R$ is then found from the contributions of both resolved and unresolved halos. A fixed simulation parameter $\Mmin$ controls the minimum considered mass of collapsing region, since small dark matter halos are generally considered to have very low star formation rates (see \mcitealt{Barkana2001} for a review) and can be ignored as not contributing a significant number of ionizing photons. 

The ionization fraction field $x_{\mathrm{HII}}(\vec{r})$ is found by determining whether the collapsed matter in a region generates enough ionizing photons to ionize the enclosed hydrogen atoms. An ionizing efficiency parameter $\Zion$ specifies how many ionizing photons are sourced per unit of collapsed matter. Pixels are painted as fully ionized if $\fcoll(\vec{r}, R) \geq \Zion^{-1}$, otherwise they are set as partially ionized according to the collapsed fraction in the cell $\Zion \fcoll(\vec{r}, R)$. Finally, \eqnref{eqn:deltaTb approx} is used to find the 21cm brightness temperature field $\deltatb (\vec{r})$ from the non-linear density field $\delta(\vec{r})$ and the neutral fraction field $x_{\mathrm{HI}}(\vec{r}) = 1 - x_{\mathrm{HII}}(\vec{r})$.

Three simulation parameters stand out as the most powerful ways to constrain reionization scenarios from data:

\begin{enumerate}
\item The ionization efficiency $\Zion$, specifying how many ionising photons are sourced per unit of collapsed matter;
\item The maximum bubble size $\Rmax$, specifying the maximum travel distance for ionizing photons from their sources;
\item The lower mass limit $\Mmin$, specifying the minimum mass of collapsed matter which produces ionizing photons.
\end{enumerate}  

\simfast{} also has the option to account for local fluctuations in the spin temperature, at the expense of considerably more computation time. We turn off this functionality to give a usable training dataset size in a reasonable time frame.

\section{Machine learning techniques}
\label{sec:ML}

The machine learning techniques in this paper are methods of multi-dimensional regression: learning the behaviour of some function $f(\vec{x})$ from noisy example training data $y_n =f(\vec{x_n}) +\text{Noise}$. The noise in all our data is sample variance from randomly seeding different density fields at the start of each simulation. \changes{We do not include instrumental noise because our emulators are intended as efficient replacements for the expensive simulations themselves. For comparison with observed telescope data, instrumental noise can be added in the comparison stage after running the clean emulated simulations.} After fitting, the models can make predicted evaluations $f(\vec{x^*})$ at new input values $\vec{x^*}$. This section describes the different machine learning techniques we used along with theoretical descriptions of their specific training methodologies. Each method learns the behaviour of the \simfast{} power spectrum for any reionization scenario specified by a continuous range of \simfast{} input parameters. The trained models can then make fast power spectrum predictions for new scenarios, provided the new scenario parameters do not lie far outside the range of our representative training data.

\subsection{Interpolation}
\label{sec:method_interpolate}
The simplest method for prediction is to interpolate the power spectrum outputs within the training data. We use two interpolation methods, linear interpolation and nearest-neighbour interpolation, implemented using the classes \textit{LinearNDInterpolator} and \textit{NearestNDInterpolator} from the \textit{scipy} module \mcitep{scipy}. These methods involve no hyperparameter searching and ignore the effect of sample variance noise in the training data. We include them as a naive benchmark to compare the accuracy and speed performance with the other models. The scipy \textit{LinearNDInterpolator} class uses \textit{qhull} from \mcite{qhull} to triangulate the input data, computing five-dimensional surfaces in the input space and then performing linear interpolation on these triangles. This process takes a long time, both for training and prediction. The scipy \textit{NearestNDInterpolator} class makes predictions by returning the output value from the nearest training data point. This process is very fast but generally results in poorer predictions.

\subsection{Multilayer perceptron}
\label{sec:mlp_training_theory}

An artificial neural network (ANN) represents the function $f(\vec{x_i})$ by manipulating its input values $\vec{x_i}$ through a series of weighted summations and simple function evaluations. This series of repeated operations can be thought of as occurring in a series of layers. The values in the first layer $\vec{h^{(0)}}$ are simply the input values $\vec{x_i}$. The network manipulates the values from one layer $h^{(l-1)}_j$ to the values in the next layer $h^{(l)}_j$ using

\begin{equation}
  \begin{array}{l}
\label{eqn:ANN}
\vec{h^{(l)}} = h^{(l)}_j = \phi_\theta \left(\ \mathlarger{\sum \limits_{i=1}^{N_i}} W^{(l)}_{ij} h^{(l-1)}_{j} \right)\,.
\end{array}
\end{equation}

\noindent The values in the $l$-th layer are a weighted sum over the values in the previous layer, using trainable weight values $W^{(l)}_{ij}$, and are then passed through an activation function $\phi_\theta(x)$. The final layer contains the network's fitted evaluations of the function, $f(\vec{x_i})$. Training the network requires finding the weight values $W^{(l)}_{ij}$ which most closely mimic the function's behaviour. 

\begin{figure}
\includegraphics[width=\columnwidth]{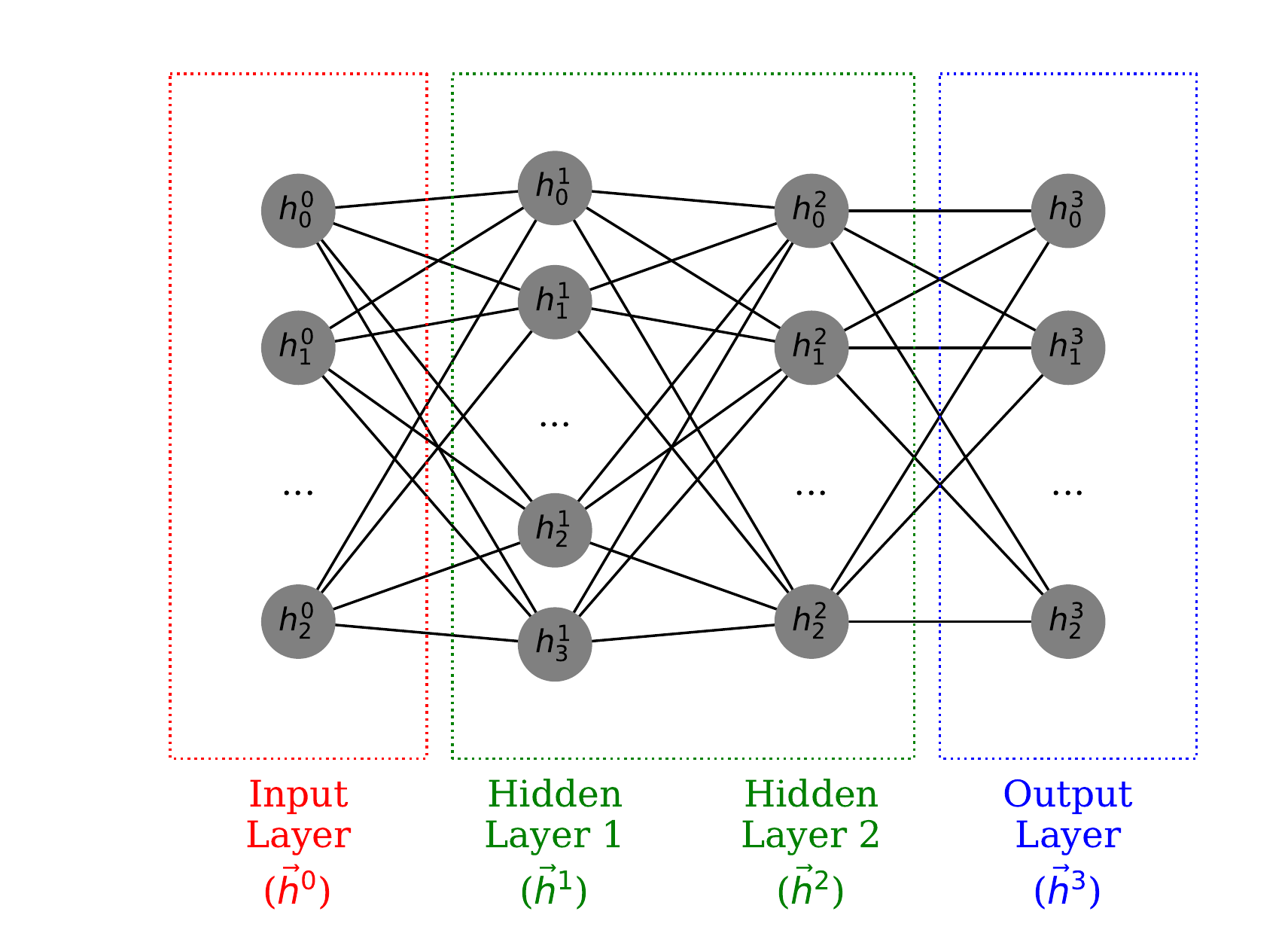}
\caption{Visualization of a multilayer perceptron with two hidden layers. Lines are weighted connections directed from left to right. Circles are the neurons which schematically hold the values, pass the weighted sum of inputs through the activation function, and send this final value to the next layer.}
\label{fig:mlp}
\end{figure}

Multilayer perceptrons (MLPs) are ANNs which contain at least one hidden layer and have a non-linear activation function. \figref{fig:mlp} shows a schematic of a typical MLP's layer structure. Lines represent the weighted connections between values. Circles represent the neurons which schematically hold the values $h^{(l)}_j$ and pass the weighted inputs through the activation function. We use the \textit{scikit-learn} package from \mcite{scikit-learn} for all our MLPs\changes{, using the following default inputs: a constant learning rate of $0.001$; batches of size $200$; the rectified linear unit function (`relu') as our activation function.}

MLP training involves finding the weight values $W^{(l)}_{ij}$ which minimize the objective function,

\begin{equation}
\label{eqn:mlp-objective}
\text{MLP Objective} =  \frac{1}{2 N}\sum \limits_{n=1}^{N} \big( f(\vec{x_n}) - y_n \big)^2  - \frac{\alpha}{2} \sum \limits_{i,j,l} \left(W^{(l)}_{ij}\right)^2
\end{equation}

\noindent for training data $(\vec{x_n},y_n)$. \changes{The weights are initialised using a different random seed for each model.} The function evaluation $f(\vec{x_n})$ in \eqnref{eqn:mlp-objective} follows the procedure given in the previous subsection: passing the input values $\vec{x_n}$ through multiple layers of weighted sums and activation function evaluations. Before training, one must fix the number of hidden layers and the number of neurons in each hidden layer. We use a fixed L2 regularization parameter \changes{value of $\alpha = 0.0001$ to reduce the effect of overfitting}. The \textit{scikit-learn} class for MLP uses backpropagation algorithm \mcite{backprop} for efficient calculation of the gradient of the objective function, see \mcite{Rumelhart1986} for a more detailed description of this algorithm. \changes{We use the `adam' optimization method \mcitep{Adam}} which terminates when the objective function falls below a tolerance of $10^{-10}$ for at least two consecutive iterations.

\subsection{Gaussian processes regression}
Gaussian process regression (GPR) is a fitting process for a function whose values are drawn from a Gaussian process. A Gaussian process is a set of random variables, any subset of which follow a jointly multi-variate Gaussian. For a finite set of $D$ random variables stored in a vector $\vec{f} = [f_1,\ldots,f_D]$, the probability density function $P(\vec{f})$ of a multi-variate Gaussian has the form

\begin{equation}
\label{eqn:GP}
\log P(\vec{f}) = - \frac{1}{2} \sum \limits _{i,j=1}^{D} \big( f_i - \mu_i \big) K_{ij} \big( f_j - \mu_j \big)\, + \mathrm{constant}.
\end{equation}

\begin{figure}
\includegraphics[width=\columnwidth]{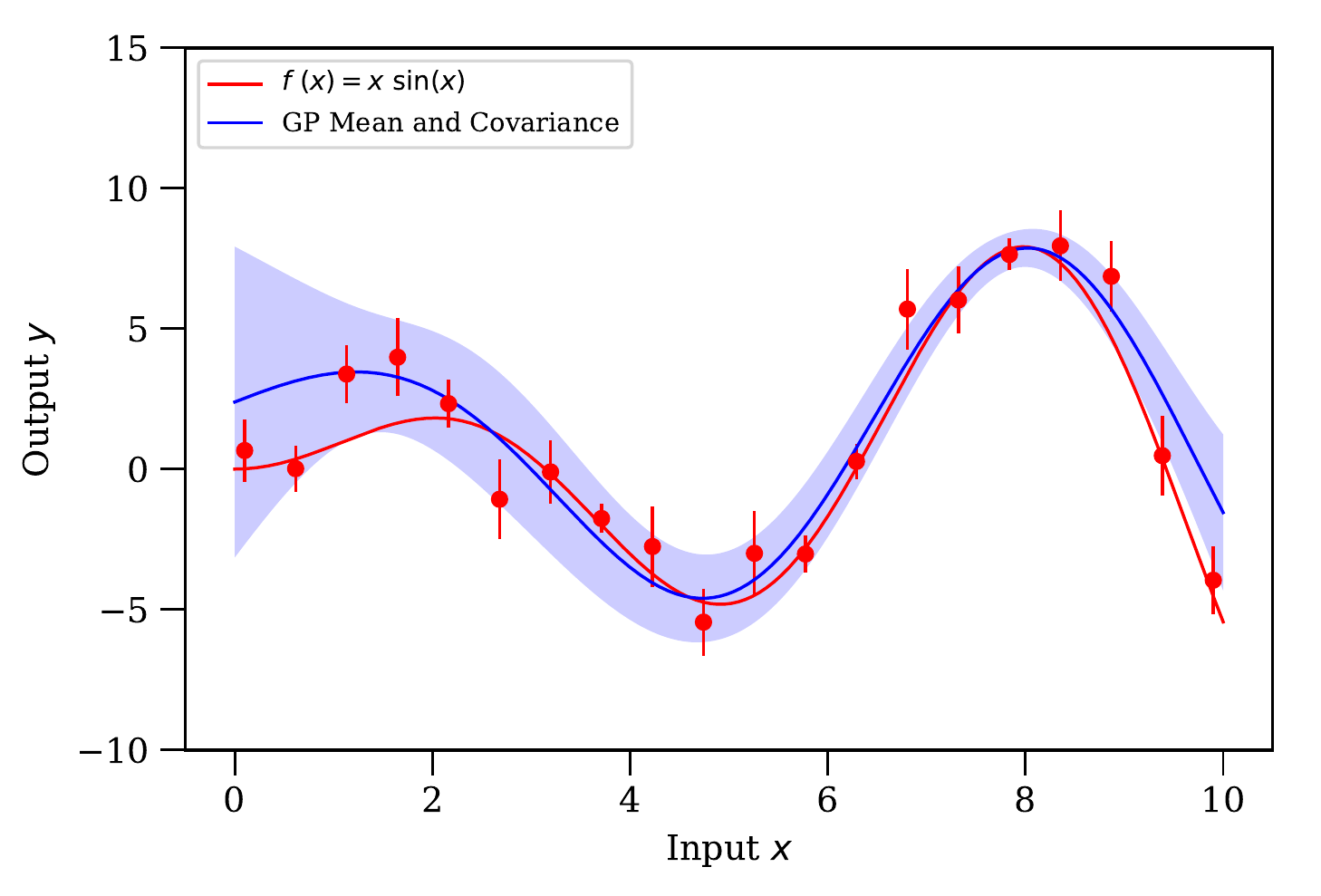}
\caption{Example of Gaussian process regression on noisy data $y_n = x_n \sin(x_n) + \mathcal{N}(0,\epsilon_n)$, with the noise amplitude on each data point $\epsilon_n$ being randomly drawn randomly from the interval $[0.5,1.5]$. The mean function (solid blue line) and covariance kernel (shaded blue region) are found which best match the training data (red points).}
\label{fig:gp_fitting}
\end{figure}

Fitting this finite distribution involves finding the elements $\vec{\mu} = [\mu_1,\ldots,\mu_D]$ of the mean vector, and the elements $K_{ij}$ of the covariance matrix. A Gaussian process extends the concept of a multi-variate Gaussian to infinite dimensions, by replacing the finite-dimensional forms [$\vec{f}$, $\vec{\mu}$, $K_{ij}$] with functional forms [$f(\vec{x})$, $m(\vec{x})$, $k(\vec{x_i},\vec{x_j})$]. A Gaussian process can then be thought of as a distribution over functions, and training involves finding the optimal forms for the mean function $m(\vec{x})$ and a covariance kernel $k(\vec{x_i},\vec{x_j})$. Predictions are made by finding the function values which maximize the joint posterior of the training data and the new input values, all of which are assumed to be drawn from the same Gaussian processes. The choice of covariance kernel reflects the expected properties of the underlying process, such as smoothness or periodicity. \figref{fig:gp_fitting} shows an example of fitting a Gaussian process, where both the fitted mean function and covariance kernel have been shown.

Gaussian process regression involves finding the likelihood distributions of the mean function $m(\vec{x})$ and covariance function $k(\vec{x_i},\vec{x_j})$ which result analytically from the noisy training data. These likelihood distributions are combined with input prior distributions, to give the final posterior distributions from which predictions can be made. Our prior for the mean function is 

\begin{equation} \label{eqn:gpflow_mean_fn}
m(\vec{x}) = \vec{A} + b \vec{x}
\end{equation}

\noindent with trainable parameters $\vec{A}$ and $b$ (initialised to zeros) specifying a linear relationship to each of the five input dimensions. Our prior for the covariance function is the Matern32 kernel, 

\begin{equation} \label{eqn:matern32}
k_{\mathrm{M32}}(\vec{x_i},\vec{x_j}) = \sigma^2 \left( 1 + \frac{\sqrt{3} |\vec{x_i}-\vec{x_j}|}{\rho} \right) \exp \left( - \frac{\sqrt{3} |\vec{x_i}-\vec{x_j}|}{\rho} \right)\,
\end{equation}

\noindent with trainable parameters for the kernel variance $\sigma^2$ and kernel length-scale $\rho$ (both initialised to unity). The Matern32 is used to represent data with a moderate level of smoothing. \changes{Both of these kernel parameters control over-fitting of this model. For instance, a smaller value of $\rho$ allows the mean function to change more rapidly as a function of the inputs, which can cause the model to overfit the training data.}

Training this model involves finding the matrix elements $K_{ij} = k(\vec{x_i},\vec{x_j})$ of the training data. The expected mean and variance for a new prediction test location $\vec{x^*}$ are then given by

\begin{equation}
\label{eqn:gpr-mean-predict}
f(\vec{x^*}) = \sum \limits_{i,j=1}^{N} k(\vec{x_i},\vec{x^*}) \big( K_{ij} + \sigma^2 \delta_{ij} \big) ^{-1} y_j\,,
\end{equation}

\begin{equation}
\label{eqn:gpr-covar-predict}
\text{Var}(f(\vec{x^*})) =  k(\vec{x^*},\vec{x^*}) - \sum \limits_{i,j=1}^{N} k(\vec{x_i},\vec{x^*}) \big( K_{ij} + \sigma^2 \delta_{ij} \big) ^{-1} k(\vec{x^*},\vec{x_j})\,,
\end{equation}

\noindent from \mcite{Rasmussen}. Note that these equations involve inverting the large matrix $(K_{ij} + \sigma^2 \delta_{ij})$, which in our case has $91000^2$ elements. Using python 8-byte \textit{float64} values, simply storing a single object instance of this matrix takes 60GB of RAM. Our computer architecture with 128GB of RAM is not large enough to invert such a matrix, since inversion requires much more RAM than a single matrix instance. Sparse Gaussian process regression (SGPR) is an approximation of GPR for huge data sets. SGPR approximates the matrix inversion by using only a subset of $m$ observed data points and inverting this smaller matrix instead. These $m$ `inducing points' are effectively an additional set of fitting parameters. Our SGPR model uses the \textit{gpflow} package\footnote{http://gpflow.readthedocs.io/en/latest/intro.html} which implements the methods in \mcite{Titsias09} using TensorFlow (\mcitealt{tensorflow2015-whitepaper}). The \textit{gpflow} package uses the \textit{scipy.optimize.minimize} function with the Limited-memory Broyden-Fletcher-Goldfarb-Shanno (L-BFGS-B) method to find the best set of inducing points. The minimization method uses the default termination method, i.e. when the maximum component of the objective function's gradient falls below a tolerance of $10^{-5}$.

\subsection{Support vector machine}
Support vector machine (SVM) models are often used for classification, but can also be used for regression. In SVM classification, training  involves finding a set of hyperplanes which separate the training data into their labelled classes while at the same time maximizing the distance between the hyperplanes and the nearest training data points. SVM regression extends this concept to functional forms, so that training the model involves finding a function $f(\vec{x})$ whose evaluations at the training points $\vec{x_n}$ are most similar to the observed training values $y_n$, while at the same time ensuring that the function is as simple as possible. We use the \textit{scikit-learn} package (\mcitealt{scikit-learn}) for our support vector machine models.

SVM training involves finding the functional form $f(\vec{x})$ such that the residual errors between the training data $(\vec{x_n},y_n)$ and the function evaluations $f(\vec{x_n})$ all lie within some tolerance $ - \epsilon \leq f(\vec{x_n}) - y_n \leq \epsilon$. This stringent constraint usually makes it impossible to find any such form $f(\vec{x})$. To weaken the condition and allow a solution, the slack variables $(\xi_n,\xi_n^*)$ are introduced so that the residual fitting error $f(\vec{x_n}) - y_n$ for the training point $(\vec{x_n},y_n)$ obeys $- \epsilon - \xi_n^* \leq f(\vec{x_n}) - y_n \leq \epsilon + \xi_n$. This optimization problem is more easily solved in the dual form, with objective function 

\begin{equation}
\label{eqn:svr_dual}
\begin{split}
\text{SVR Objective} = \ & \sum \limits_{i,j=1}^{N}  \big(\alpha_i - \alpha_i^*\big)\ k(\vec{x_i},\vec{x_j})\ \big(\alpha_j - \alpha_j^*\big) \\ 
& + \epsilon \sum \limits_{i=1}^{N} \big(\alpha_i + \alpha_i^*\big) - \sum \limits_{i=1}^{N} \Big( y_i (\alpha_i - \alpha_i^*) \Big)\,.
\end{split}
\end{equation}

\noindent Training involves finding the values $(\alpha_i, \alpha_i^*)$ which minimize this objective function, subject to margin constraints

\begin{equation}
\begin{split}
\label{eqn:svr_constraints}
& \sum \limits_{i=1}^{N} \big(\alpha_i - \alpha_i^*\big) = 0\,\ \ \text{and}\ \,  0 \leq \alpha_i, \alpha_i^* \leq C\,.
\end{split}
\end{equation}

The kernel function $k(\vec{x_i},\vec{x_j})$ in \eqnref{eqn:svr_dual} controls the functional form $f(\vec{x})$. We try three different kernel functions: radial basis function (RBF), polynomial, and sigmoid. As discussed in \secref{sec:svm-hyper} later, the only kernel which gives rise to reasonable accuracy predictions (with MSE less than 500 \%) is the RBF kernel,

\begin{equation} \label{eqn:RBF_kernel}
k_{\mathrm{RBF}}(\vec{x_i}, \vec{x_j}) = \exp \big( - \gamma \left| \vec{x_i}-\vec{x_j} \right|^2 \big)\,.
\end{equation}

\noindent The RBF kernel is infinitely differentiable, hence is often used to model data from smooth distributions. Before training, one must set the penalty term $C$, the kernel influence range $\gamma$ (hereafter written $\textit{gamma}$ to match the python class parameter), and the margin tolerance $\epsilon$ (written $\textit{epsilon}$). \changes{Overfitting for SVR models is discouraged by C the penalty term.}

\section{Emulator training}
\label{sec:training}
In this section we describe how we create the training data and the specific choices we make in training our emulators. Standard practice is to use a large training dataset and then check that the trained emulators make valid predictions for unseen validation data. The training results are given later in \secref{sec:training results}. All emulators are trained on the same architecture, each on a single node using 16 Xeon E5-2650 cores and 128GB RAM.

\subsection{\simfast{} simulations}
\label{sec:training_set_design_simfast21}
We run 2000 \simfast{} simulations in total, retaining only the three input reionization parameters and the final output spherically averaged power spectra for each simulation. We use 1000 simulations for training, 500 for validation, and another 500 for testing the emulators which have highest prediction accuracy on the validation data. Each simulation generates three-dimensional realisations of the $\deltatb$ field in a cube of size $500\text{Mpc}$ resolved into $512^3$ pixels (smoothed from density fields resolved into $1536^3$ pixels). \changes{This gives power spectra values for seven redshift values: $\{8.0, 9.5, 11.0, 12.5, 14.0, 15.5, 17.0\}$} and thirteen k-values in the range $\{0.02, 3.0\}\ \kunit$. This corresponds to $91000$ overall training data points, and $45500$ data points each for validation and testing. The power spectra data have size of 335MB for all 2000 simulations, compared to 7TB size of all $\deltatb$ boxes.

\subsection{Training set design}
Our emulators map five input values to a single output target value. The target value is the $\deltatb$ power spectrum value for the given inputs. The first three input values are the three reionization parameters (see \secref{sec:simfast21}), which are different for each simulation. The final two inputs are the redshift $z$ and the $k$-value, the values for which are constant across all simulations and are given in \secref{sec:training_set_design_simfast21}. The function $f(\vec{x})$ which the models are fitting is then the spherically averaged 21cm power spectrum $P_{\Deltatb}(\Mmin, \Zion, \Rmax, z, k)$.

We use the Latin Hypercube method designed by \mcite{LatinHypercube} to choose the reionization parameter values for our simulations. The Latin Hypercube method provides a way to sample the three-dimensional input space in a more efficient way than naive exhaustive grid-search. We use the following ranges and scalings for the reionization parameters:

\begin{enumerate}
\item $\Mmin$ in the logarithmic range $[10^{7.8}, 10^{9.8}]$
\item $\Zion$ in the linear range $[5, 100]$
\item $\Rmax$ in the linear range $[5, 20]$ 
\end{enumerate}

\noindent \changes{These ranges match those used by the semi-numerical simulation authors, see for example \mcite{21CMMC2015}. The lower $\Mmin$ limit comes from the lowest temperature at which atomic hydrogen can cool and accrete onto halos, and the upper limit from observations of high-redshift Lyman break galaxies \changes{\citep{21CMMC2015}}. The $\Zion$ range roughly corresponds to ionizing photon escape fractions of 5\% to 100\%. The $\Rmax$ range arises from recombination models of \mcite{Sobacchi&Mesinger2014}, and only has an effect near the end of reionization when the ionized bubble sizes are comparable to $\Rmax$ (\mcitealt{Alvarez&Abel2012}, \mcitealt{McQuinn2007}). See Figures \ref{fig:power_spectra_predict_ion-eff} and \ref{fig:power_spectra_predict_mmin} later for example power spectra across these ranges for $\Zion$ and $\Mmin$ values.}

We also test three different scaling types for the target values to determine which gives the most accurate emulation. These three are a linear function $y = P_k$, a logarithmic function $y = \log[P_k]$, and a pseudo-logarithmic function $y = \sinh^{-1}[P_k]$ sometimes called luptitude after \mcite{luptitude}. We test logarithmic scaling as an attempt to exploit the fact that power spectra appear more naturally spaced in logarithmic space $\log[P_k]$ than in linear space $P_k$. However a few percent of the power spectra data are zero-valued, especially at early and late redshifts where the ionization field $x_{\mathrm{HII}}(\vec{r})$ becomes uniform and $\deltatb$ is effectively zero everywhere (\mcitealt{PritchardLoeb2012}, pages 12-13). Our motivation for luptitude scaling is to retain as much data as possible: a purely logarithmic scaling would require us to throw away all zero-valued data points and reduce the size of our training data set. We comment on the effects of including or excluding these zero-valued data in \secref{sec:low-z-issues}.

\subsection{k-range restriction}
\label{sec:k-range-restrictions}
We exclude the largest and smallest scales from our validation and testing data, including only $0.1 \leq k \leq 2.0$ values. On large scales ($k<0.1\ \kunit$), the power spectrum is affected by foregrounds \mcite{Datta2010}. The finite resolution of our simulations means that there is little information in the power spectrum on very small scales ($k>2.0\ \kunit$). These restrictions are common for semi-numerical simulations, see for example \mcite{21CMMC2015}.

\subsection{Goodness of fit evaluations}
\label{sec:goodness_of_fit}
For validation and testing, we measure the goodness of fit between predicted target values $y^*(k,z)$ and measured target values $y(k,z)$ using the mean squared error

\begin{equation}
\label{eqn:mse}
\text{MSE} \big[ y(k,z),y^*(k,z) \big] = \frac{1}{N_{\mathrm{z}}N_{\mathrm{k}}} \sum \limits_{z}^{N_{\mathrm{z}}} \sum \limits_{k}^{N_{\mathrm{k}}}  \left( \frac{y(k,z) - y^*(k,z)}{y(k,z)} \right)^2\
\end{equation}

\noindent along with the percentage mean squared error, $100 \times \text{MSE}$. The MSE is averaged over all $N_{\mathrm{z}}$ redshifts values and all $N_{\mathrm{k}}$ scale values in the range $0.1 \leq k \leq 2.0$, unless explicitly mentioned otherwise. For comparability, we use this same error function for all different emulators during validation and testing, although the models use different error metrics for determining their training convergence (see \secref{sec:ML} for the training objective functions for each model).

\section{Emulator training results}
\label{sec:training results}
After training each emulator, we test its accuracy by generating predictions for a set of unseen validation data. By calculating the MSE value in \eqnref{eqn:mse} between the predicted outputs and the true outputs, we determine which emulator makes the most accurate predictions. A low MSE means a high prediction accuracy. 

\subsection{Target value scaling}
Here we compare the prediction accuracy for the three scaling methods of the target power spectra values: linear, logarithmic, and pseudo-logarithmic $\sinh^{-1}(x)$. As expected, the linear function has poor prediction accuracy because the power spectra values are more naturally spaced in logarithmic space than in linear space. The logarithmic function works fairly well at intermediate redshifts for this same reason, but all of the zero-valued power spectrum values had to be discarded as $\log(0)$ is undefined. The pseudo-logarithmic function $\sinh^{-1}(x)$ has the highest prediction accuracy over all redshifts and allows us to retain all training data points (with zero-valued outputs or otherwise). We use the pseudo-logarithmic function in all our emulators from here on.

\subsection{Hyperparameter searching}
\label{sec:bv theory}

Each model has a set of trainable values referred to as fitting parameters. Many models have an additional set of values which must be fixed even before starting to train, referred to as hyperparameters. Here we describe which hyperparameters (if any) we vary for each model type, and which hyperparameters give rise to the best prediction accuracy. For each model, we restrict the total training time for all hyperparameter searching to 156 CPU hours. The interpolation models involve no hyperparameters, and for the SGPR model we simply increase the number of inducing points $m$ until the individual model's training time reaches 156 CPU hours. Increasing $m$ should always increase the SGPR model's accuracy and so the value of $m$ is not treated as a hyperparameter when considering the total training time. Including models with smaller $m$ values in the total training time would give a smaller maximum value of $m$, making an unfair comparison with the other models.

\subsubsection{MLP layer sizes}
\begin{figure}
\includegraphics[width=\columnwidth]{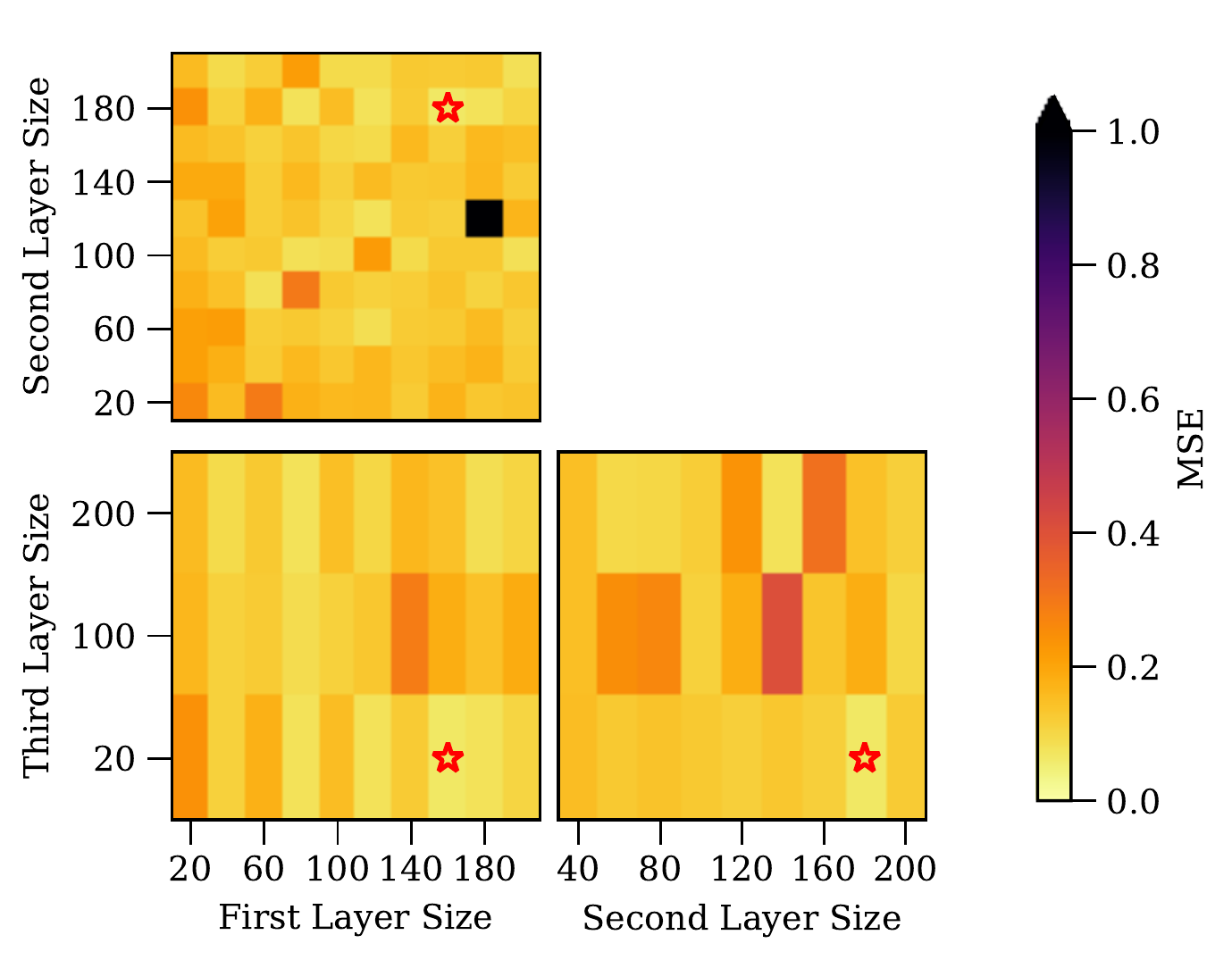}
\caption{Mean squared error on the validation data for three-layer multilayer perceptron models, as a function of the sizes of each hidden layer. The red star shows the layer sizes of the MLP emulator with the highest prediction accuracy: 160 neurons in the first hidden layer, 180 neurons in the second hidden layer, and 20 neurons in the final hidden layer.}
\label{fig:mlp_hyper_3}
\end{figure}

We use MLP models with one, two and three hidden layers. The sizes of the hidden layers were varied linearly in the range $[0, 200]$ using a simple grid-search method. Generally, the emulator models with more hidden layers have higher prediction accuracy. The validation MSE values for the best one-, two- and three-layer MLP emulators are $13 \%$, $2.3 \%$, and $1.6 \%$ respectively. The validation MSE values for three-layer MLP emulators are shown in \figref{fig:mlp_hyper_3} as a function of the sizes of each of the three hidden layers. Most three-layer MLP models have a low validation MSE near $10\%$. The best emulator has hidden layers sizes $160-180-20$, the hyperparameters for which are indicated by the location of the red star. \changes{Our MLP models end training when the objective function changes more slowly than a threshold tolerance for several training epochs. Most of our MLP models achieved this in fewer than $400$ training epochs, with some 1-layer models lasting up to $800$ epochs.}

\begin{figure}
\includegraphics[width=\columnwidth]{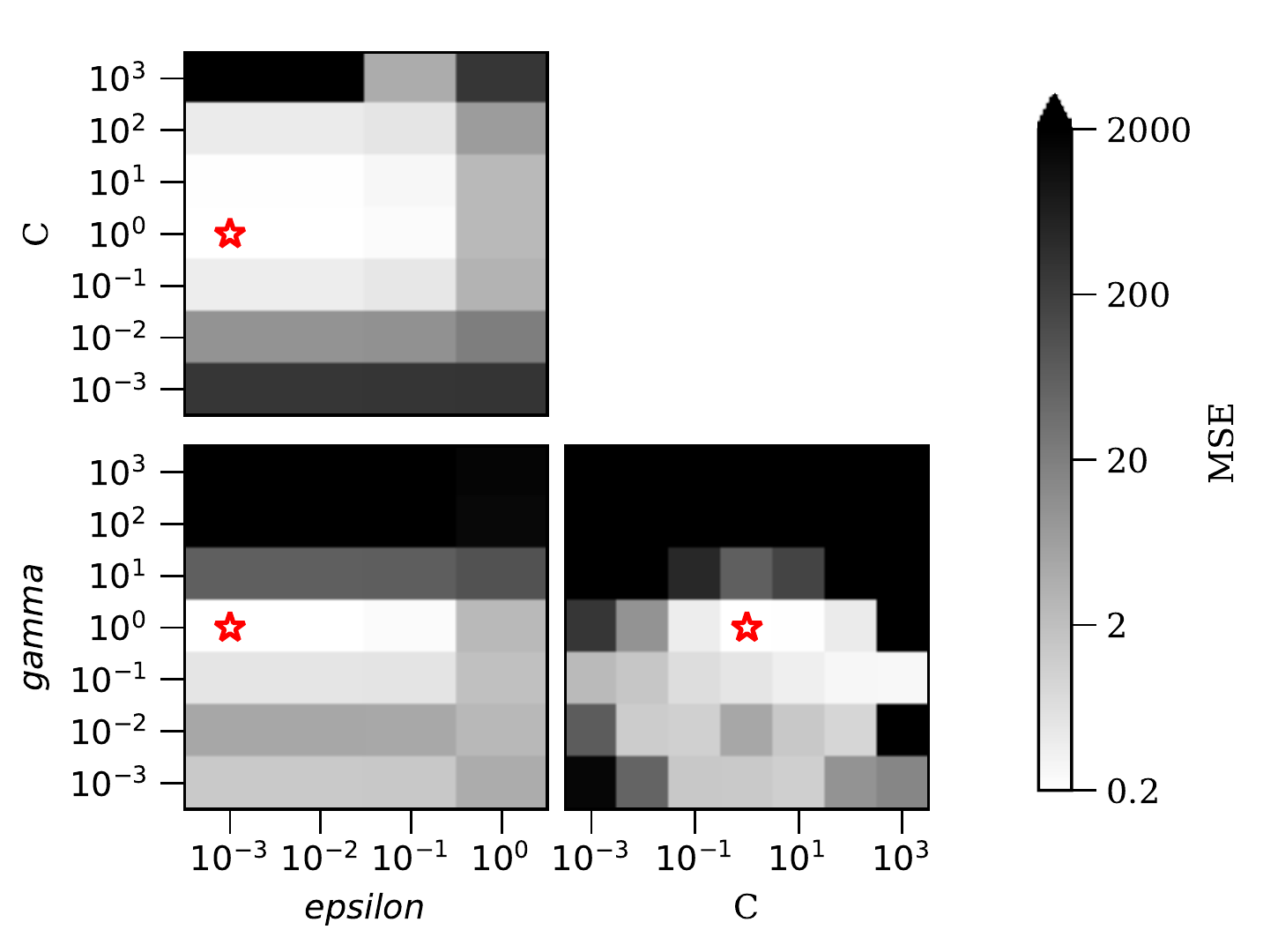}
\caption{Mean squared error on the validation data as a function of the model hyperparameters, for support vector machine emulators using the RBF kernel. The hyperparameters are the penalty term $\textit{C}$, margin tolerance \textit{epsilon} and influence range \textit{gamma}. The spread of MSE values is much larger for SVM models, indicated by the logarithmic colour scale between MSE values of $10^{-1}$ and $10^3$. The hyperparameters of the highest prediction accuracy SVM model are indicated by the red star: $\textit{C}=1.0$, $\textit{epsilon}=10^{-3}$ and $\textit{gamma} = 1.0$.}
\label{fig:svm_hyper}
\end{figure}

\subsubsection{SVM margin hyperparameters} \label{sec:svm-hyper}
We test a range of SVM emulators with different values for three hyperparameters controlling the margin. We vary the penalty parameter C logarithmically in the range $[10^{-3},10^{3}]$; the tolerance \textit{epsilon} logarithmically in the range $[10^{-3},10^{0}]$; and the kernel influence range \textit{gamma} logarithmically in the range $[10^{-3}, 10^{3}]$. These hyperparameters are the suggested ranges by \textit{sklearn} and we use a simple grid-search to find the best hyperparameters. We also test three kernel functions: RBF, sigmoid, and polynomial. \figref{fig:svm_hyper} shows how the validation MSE of emulators using the RBF kernel depends on the SVM hyperparameters. The different colour-map is used to emphasise that the colour range is logarithmic and has a much larger spread of MSE values between $0.2$ and $2000$ (or between $20 \%$ and $2 \times 10^{5} \%$). The best SVM emulator has validation MSE of $20 \%$, using hyperparameters $\textit{C}=1.0$, $\textit{epsilon} = 10^{-3}$, $\textit{gamma} = 1.0$ and the RBF kernel. All SVM emulators with kernels other than RBF have much worse validation MSE: the best polynomial and sigmoid SVM emulators have validation MSEs of $50000 \%$ and $500 \%$ respectively.

\subsection{Overfitting tests}
\label{sec:bv}

\begin{figure} 
\includegraphics[width=\columnwidth]{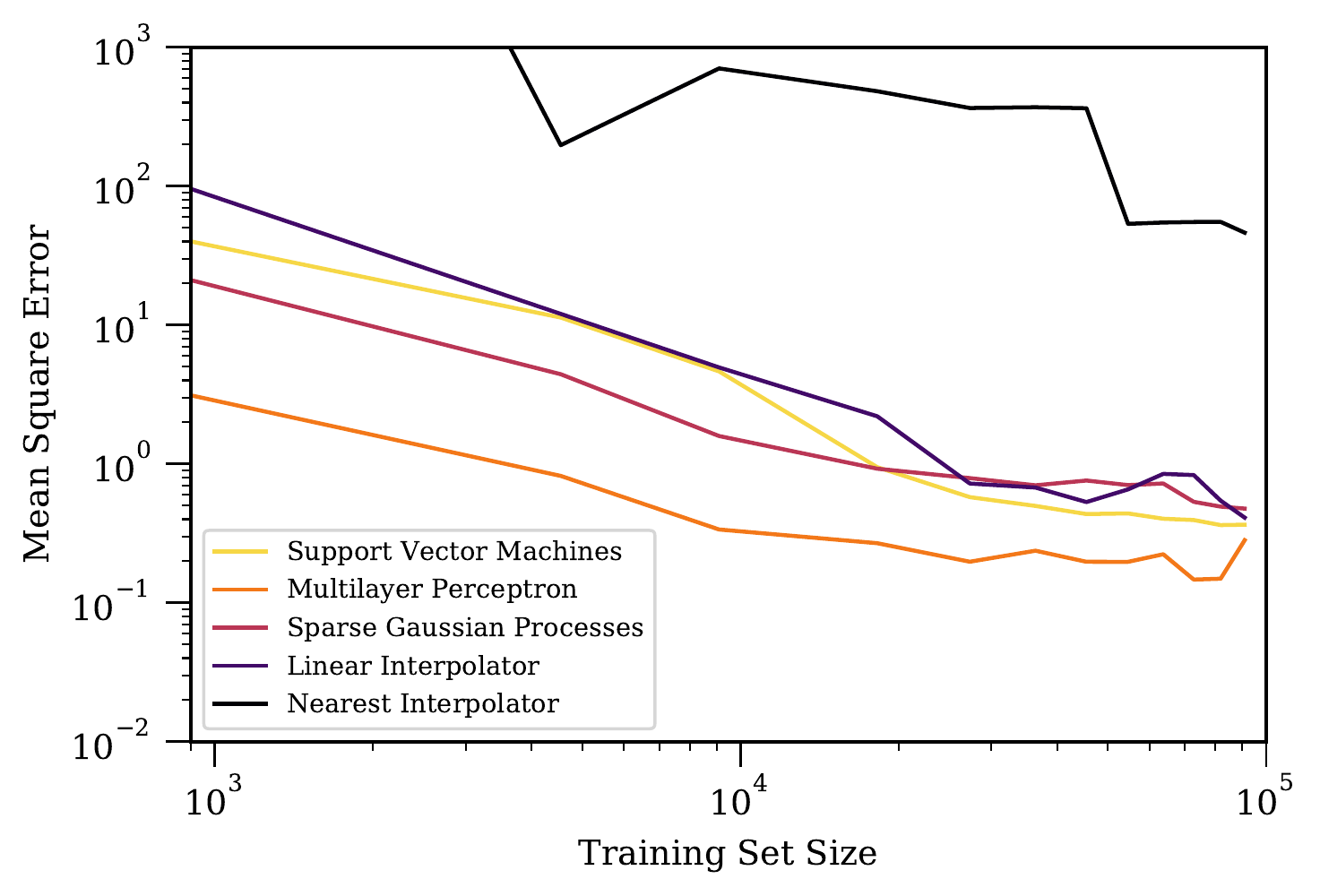}
\caption{\changes{Mean squared error on the validation data as a function of training set size. The best hyperparameters were fixed for each model, and the emulator retrained with more training information. The MSE curves generally improve with more training data, implying that none has been overfitted.}}
\label{fig:bias_variance}
\end{figure}

For each model we determine the best hyperparameters by trying a range of values and selecting the emulator which shows the highest prediction accuracy on the validation data. By trying different hyperparameter values we can usually find a closer fit to the data. However, this process is sensitive to over-fitting: the model might fit the training data more closely, but it may not extend well to new data. We test for overfitting by training a series of emulators with increasing training dataset sizes, keeping the hyperparameters fixed at the proposed best values. Providing more training data should give rise to improved predictions for the unseen validation data. If providing more training data instead leads to a decrease in validation prediction accuracy, then overfitting has occurred: the model makes good predictions for the training data, but does not extend well to new input values. \figref{fig:bias_variance} shows the results of these tests, giving the mean square error on the validation data for each model, using differently sized training datasets. \changes{All mean squared errors} generally decrease with increased training set size, implying that none has been overfitted.

\subsection{Performance on testing data}
\begin{figure}
\includegraphics[width=\columnwidth]{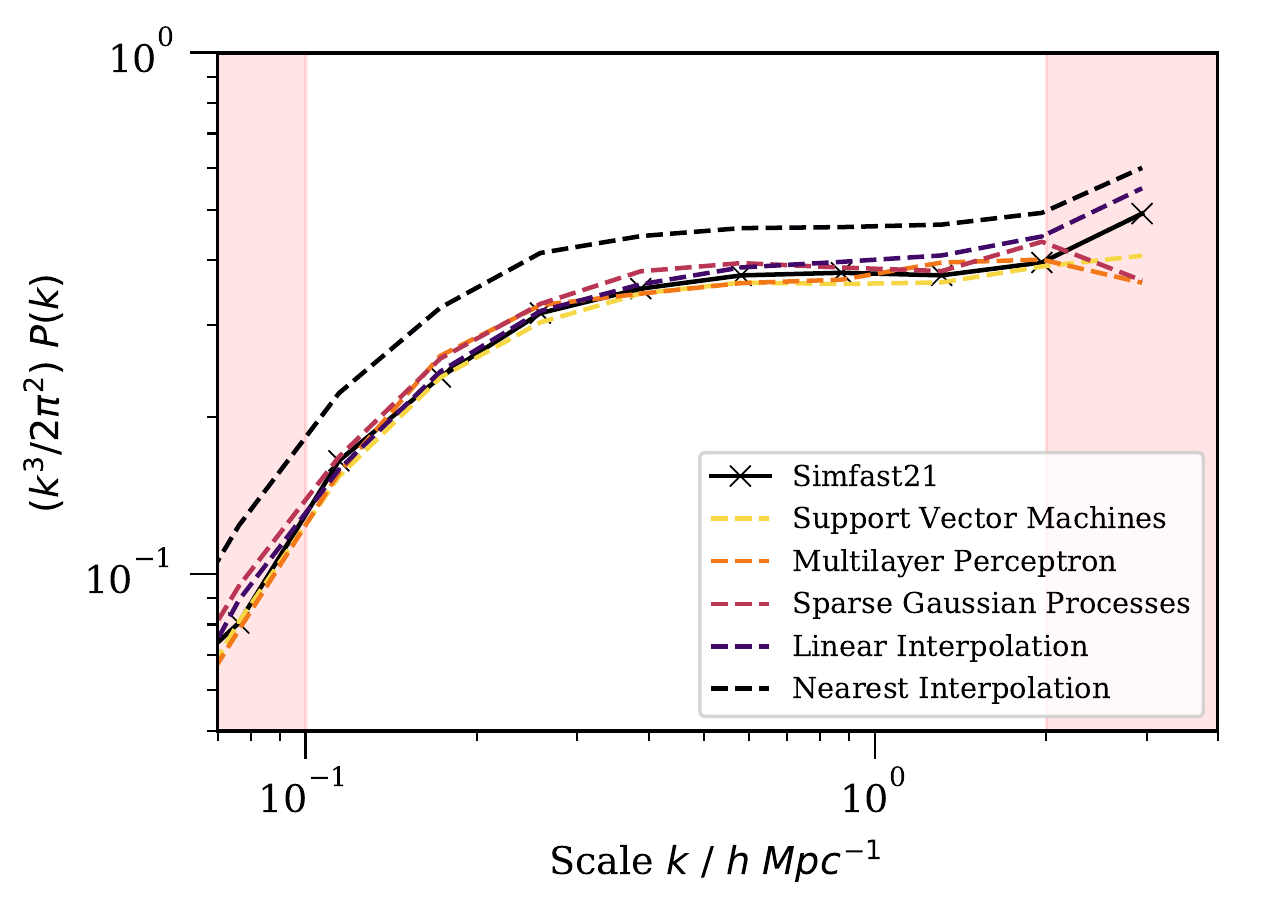}
\caption{Predicted $\deltatb$ power spectra of a canonical simulation with reionization parameters $\{5 \times 10^{8} M_{\odot}, 30.0, 10\ \text{Mpc}\}$. Dotted lines show the predictions from the best emulator of each type. Solid line shows the power spectrum from an actual \simfast{} simulation. The red shaded areas indicate the $k$-values that were excluded from our validation and testing. This test simulation was chosen from the testing data as the nearest to the canonical reionization parameters. The model using nearest-neighbour interpolation has significantly different predictions, likely owing to the underfitting processes discussed in \secref{sec:discussion-emulators-speed-accuracy}.}
\label{fig:predictions}
\end{figure}

Here we test the performance of the best emulator for each model type using all 500 simulations in our testing set. \tabref{tab:emulator_performance} shows the accuracy and speed of each emulator for making predictions on the entire testing dataset. The global MSE percentage is averaged across the entire testing data set. In \secref{sec:low-z-issues} later, we discuss the fact that our emulators have worse accuracy at lower redshifts. We include a column in \tabref{tab:emulator_performance} for the percentage MSE averaged across the testing data with higher redshifts ($z \geq 10$). \figref{fig:predictions} shows an example of the power spectra outputs from the best emulator of each model type, showing the predictions for a single test simulation near the canonical reionization parameters at $z=9.5$.

\section{Emulator training discussion}
\label{sec:discussion-emulators}

\begin{figure*}
\centering
\subfloat[Three-layer Multilayer Perceptron]{
\label{fig:best_regions_MLP3}
  \includegraphics[width=0.42\textwidth]{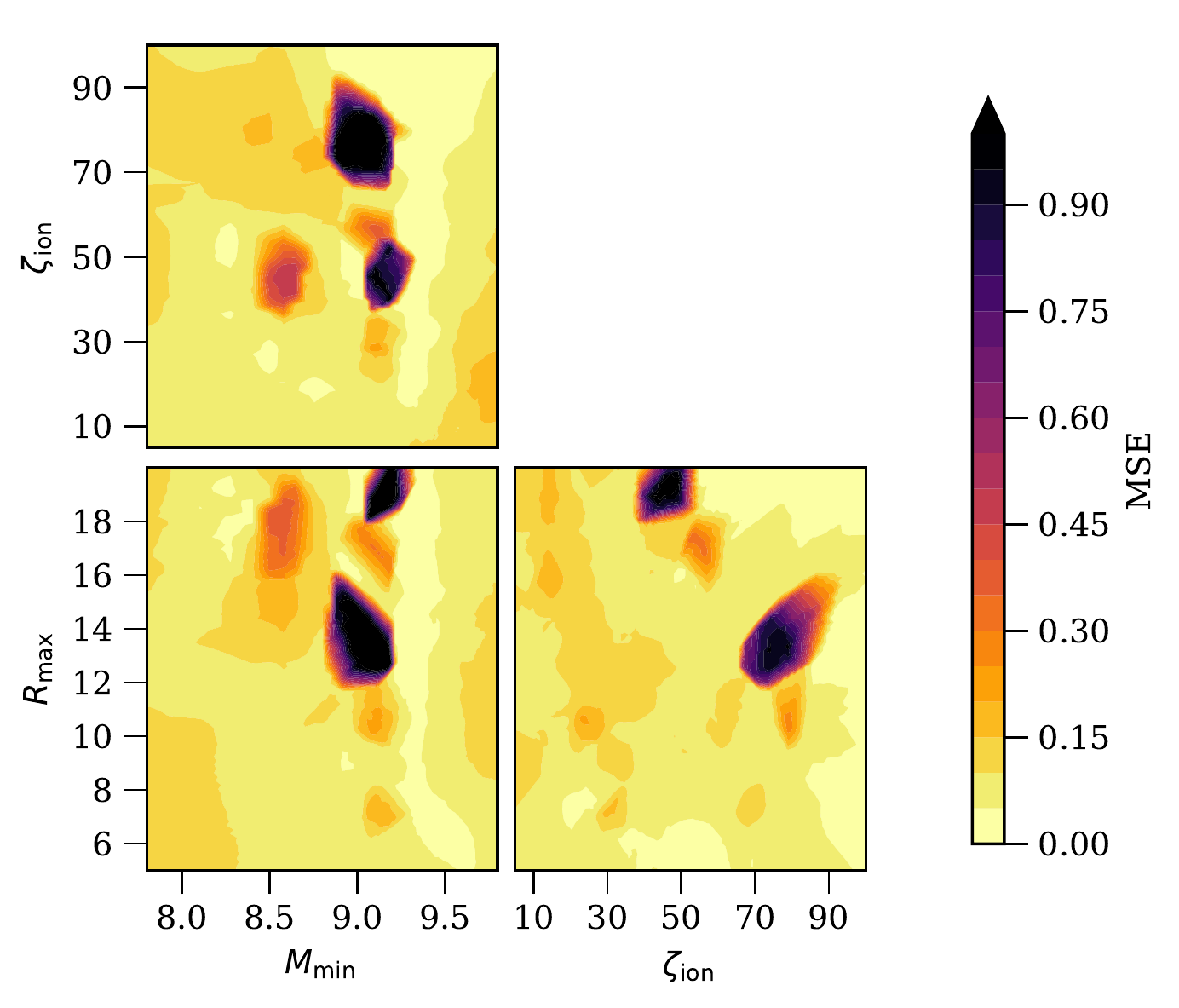}
}
\hfill
\subfloat[Nearest ND Interpolation]{
\label{fig:best_regions_NND}
  \includegraphics[width=0.42\textwidth]{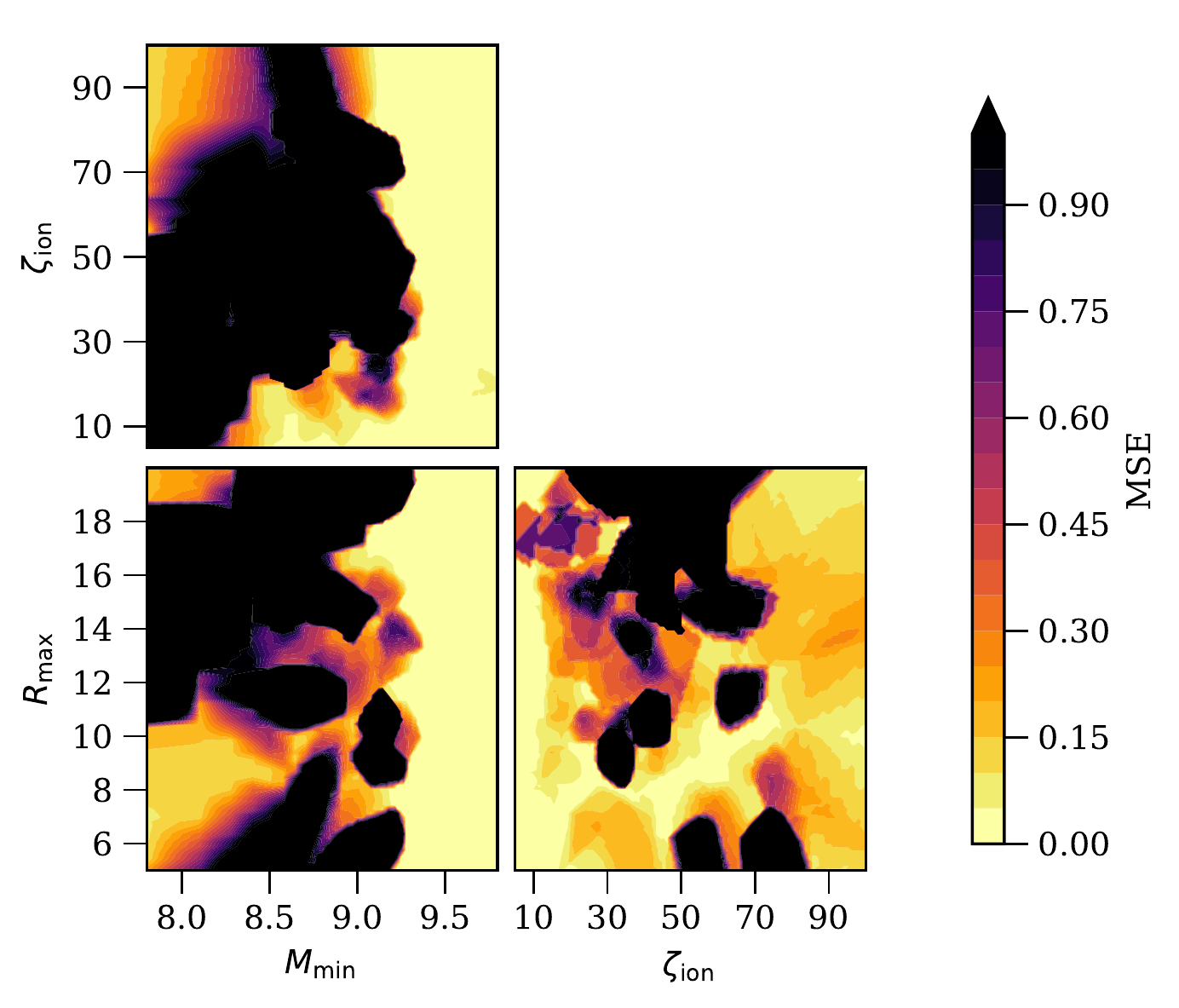}
}
\subfloat[Linear ND Interpolation]{
\label{fig:best_regions_LND}
  \includegraphics[width=0.42\textwidth]{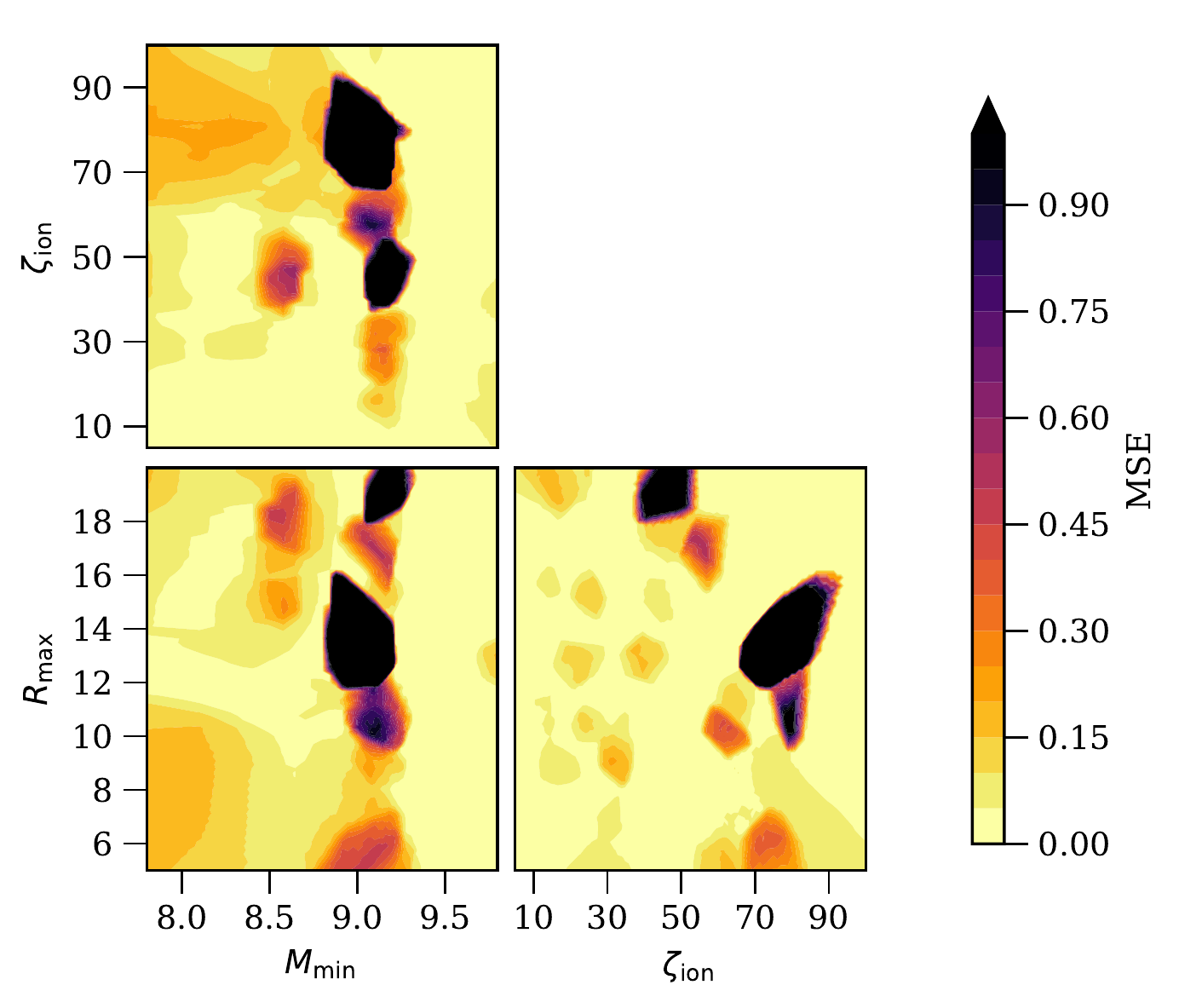}
}
\hfill
\subfloat[Sparse Gaussian Processes]{
\label{fig:best_regions_SGPR}
  \includegraphics[width=0.42\textwidth]{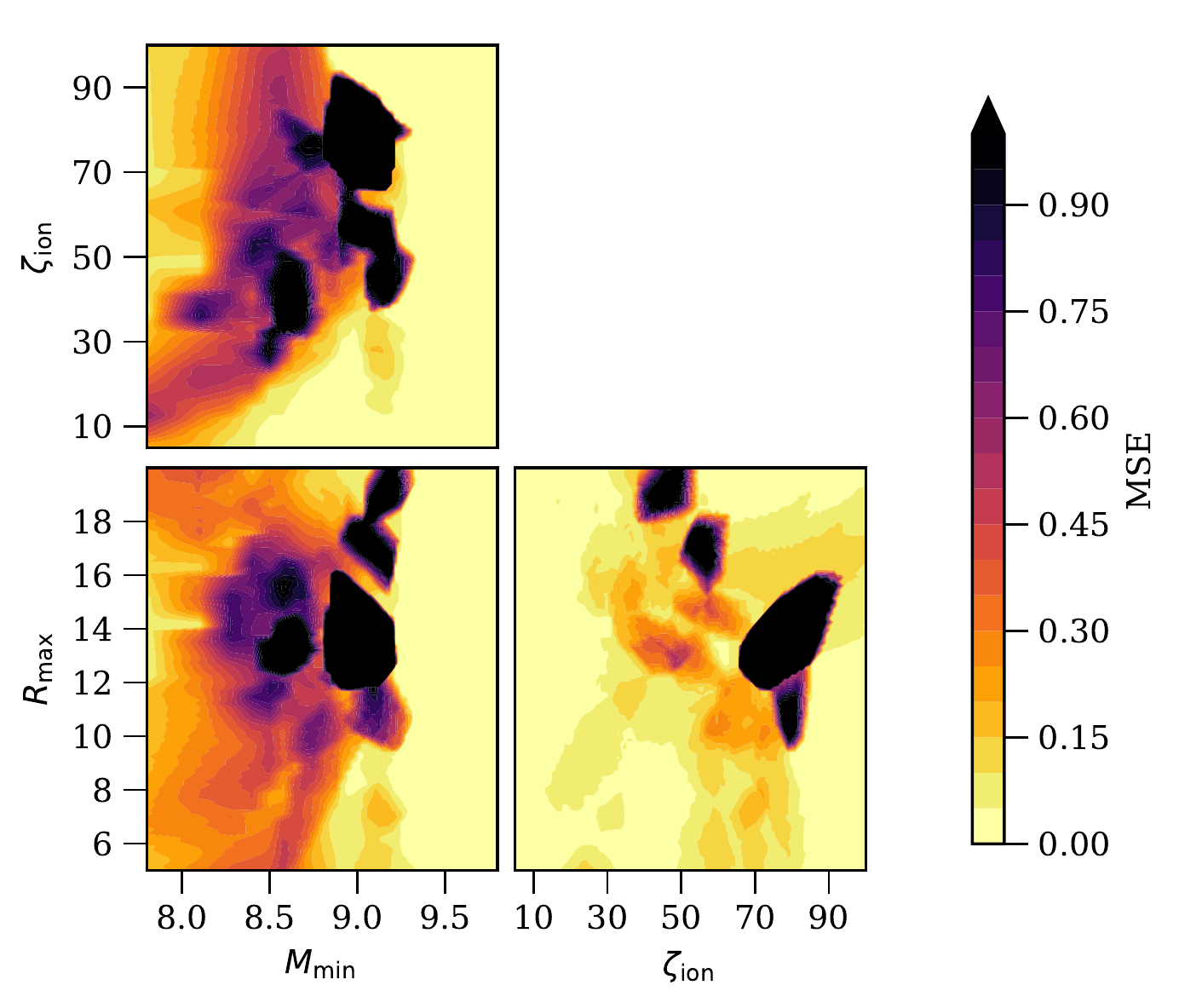}
}
\subfloat[Support Vector Machine]{
\label{fig:best_regions_SVR}
  \includegraphics[width=0.42\textwidth]{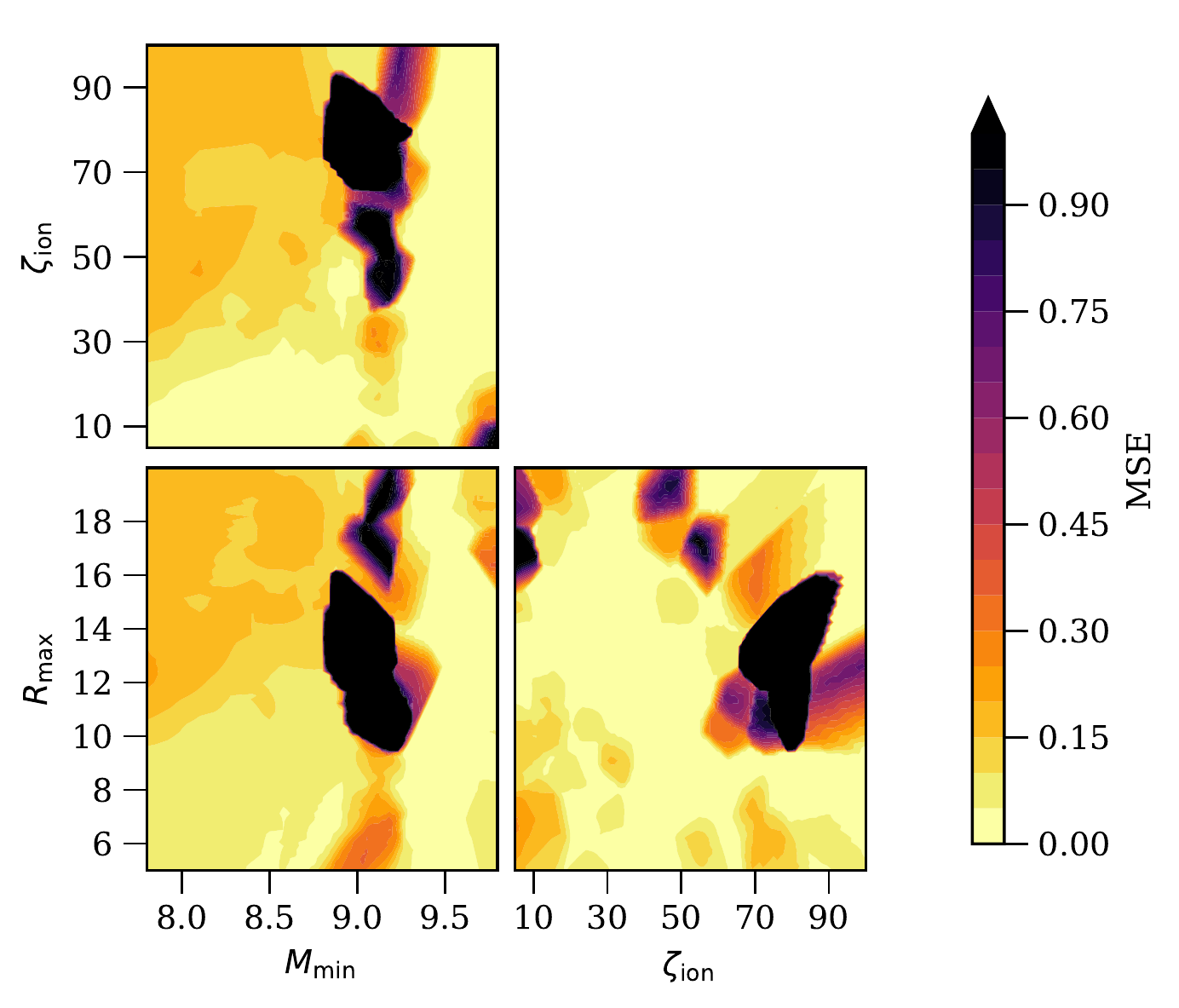}
}
\caption{Mean squared error on testing dataset for the five model types, as a function of prediction location in the three-dimensional reionization parameter space. Each panel shows the MSE values marginalized over all but two input dimensions. For instance, the $\Zion$-$\Mmin$ panels show the MSE values as marginalised over \{$\Rmax$, $z$, and $k$\} dimensions. }
\label{fig:best_regions}
\end{figure*}

\figref{fig:best_regions} shows the prediction MSE of each emulator as a function of location in parameter space. The dark regions indicate the regions of parameter space which are most difficult to emulate. All panels in \figref{fig:best_regions} show a region of poorer prediction accuracy for inputs near $\Mmin=10^9$. This is likely due to the finite mass resolution of our \simfast{} simulations. For values of $\Mmin$ near the mass resolution, the simulation switches between containing both resolved and unresolved halos (if $\Mmin < 5 \times 10^9)$, and containing only resolved halos (if $\Mmin > 5 \times 10^9)$. The change in behaviour appears to be difficult to emulate for all model types.

\begin{center}
\begin{table*}
\centering
\begin{tabular}{ l | c | c | c }
Model Type & Global Test MSE \%  (all $z$) & Test MSE \% for $z \geq 10$ & Prediction Time \\
\hline
Nearest Neighbour Interpolation & 290 \% $^{*}$ & 5.1 \% & 0.20s  \\
Sparse Gaussian Processes $m=2730$ & 36 \%& 0.6 \% & 116s \\
Support Vector Machine  & 32 \% & 2.1 \% & 27s \\
1-layer Multilayer Perceptron  & 27 \% & 9.2 \% & 0.07s \\
Linear Interpolation & 17 \% & 1.6 \% & 4.1 hours $^{*}$ \\
2-layer Multilayer Perceptron & 4.5 \% & 2.3 \% & 0.14s \\
3-layer Multilayer Perceptron & 3.8 \% & 1.4 \% & 0.27s \\
\end{tabular}
\caption{Speed and accuracy performance of the best emulator for each technique, using the testing data set. The percentage MSE values here are $100 \times \text{MSE}$. The rows are sorted in order of global prediction accuracy, from highest MSE (least accurate) at the top to lowest MSE (most accurate) at the bottom. We give global MSE values averaged across the entire dataset and also the MSE for a subset of the testing data with $z \geq 10$ to demonstrate that most of the poor accuracy occurs at later redshifts. See \secref{sec:discussion-emulators} for a discussion on the extreme ($^{*}$) values for the two naive interpolation methods. For all models except SGPR, the total time for hyperparameter searches is 156 CPU hours. For SGPR model, we run a single model with the largest possible number of inducing points $m$ without exceeding 156 hours training time.}
\label{tab:emulator_performance}
\end{table*}
\end{center}

\subsection{Speed and accuracy performance}
\label{sec:discussion-emulators-speed-accuracy}

\begin{figure}
\includegraphics[width=\columnwidth]{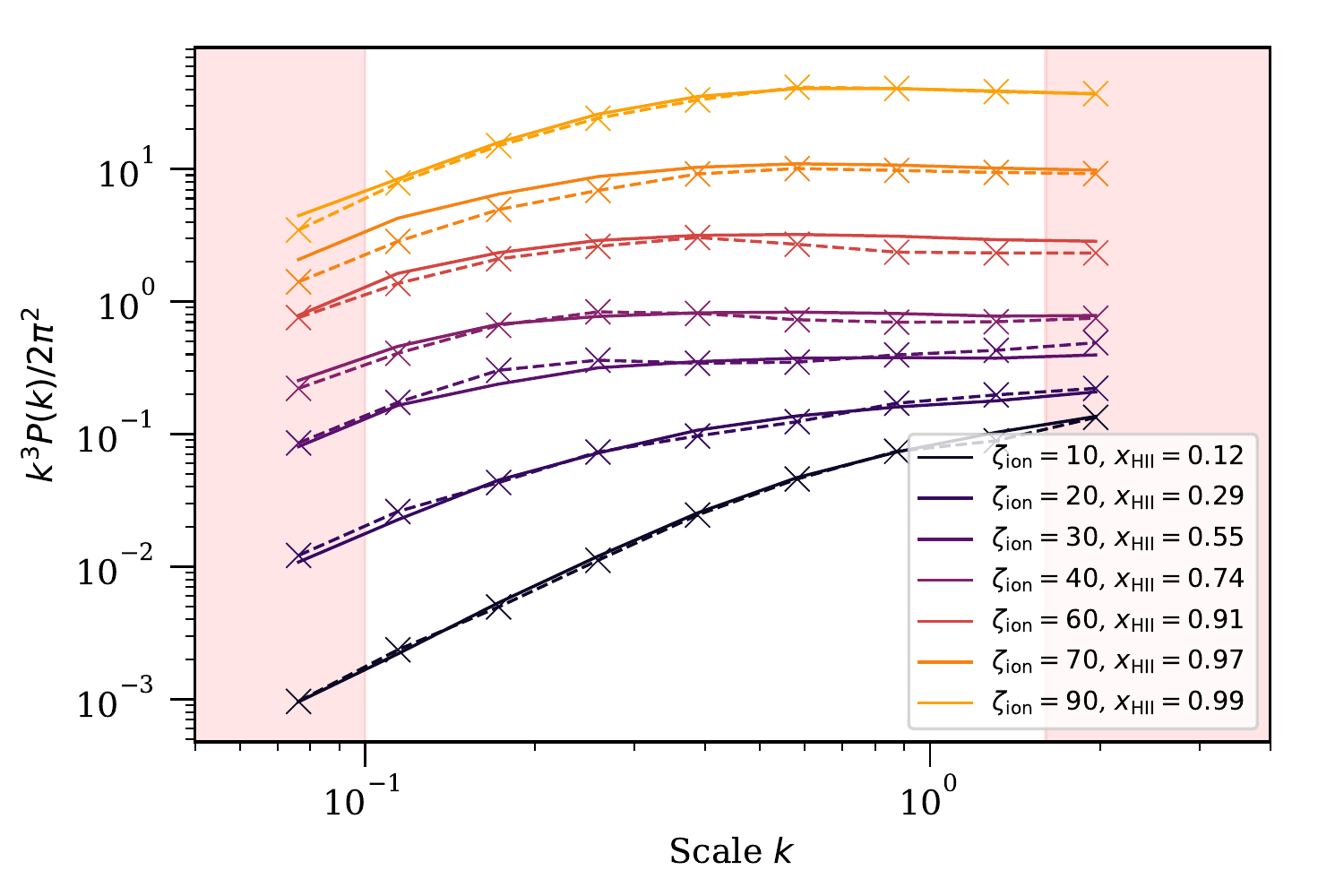}
\caption{\changes{Example emulated and simulated power spectra for a range of $\Zion$ values at $z=9.5$, for fixed $\Mmin = 5 \times 10^8$ and $\Rmax = 10$. Solid line shows the simulated power spectra, dotted line shows the predicted power spectra from our best emulator. The ionization fraction for each line is given in the legend.}}
\label{fig:power_spectra_predict_ion-eff}
\end{figure}

\begin{figure}
\includegraphics[width=\columnwidth]{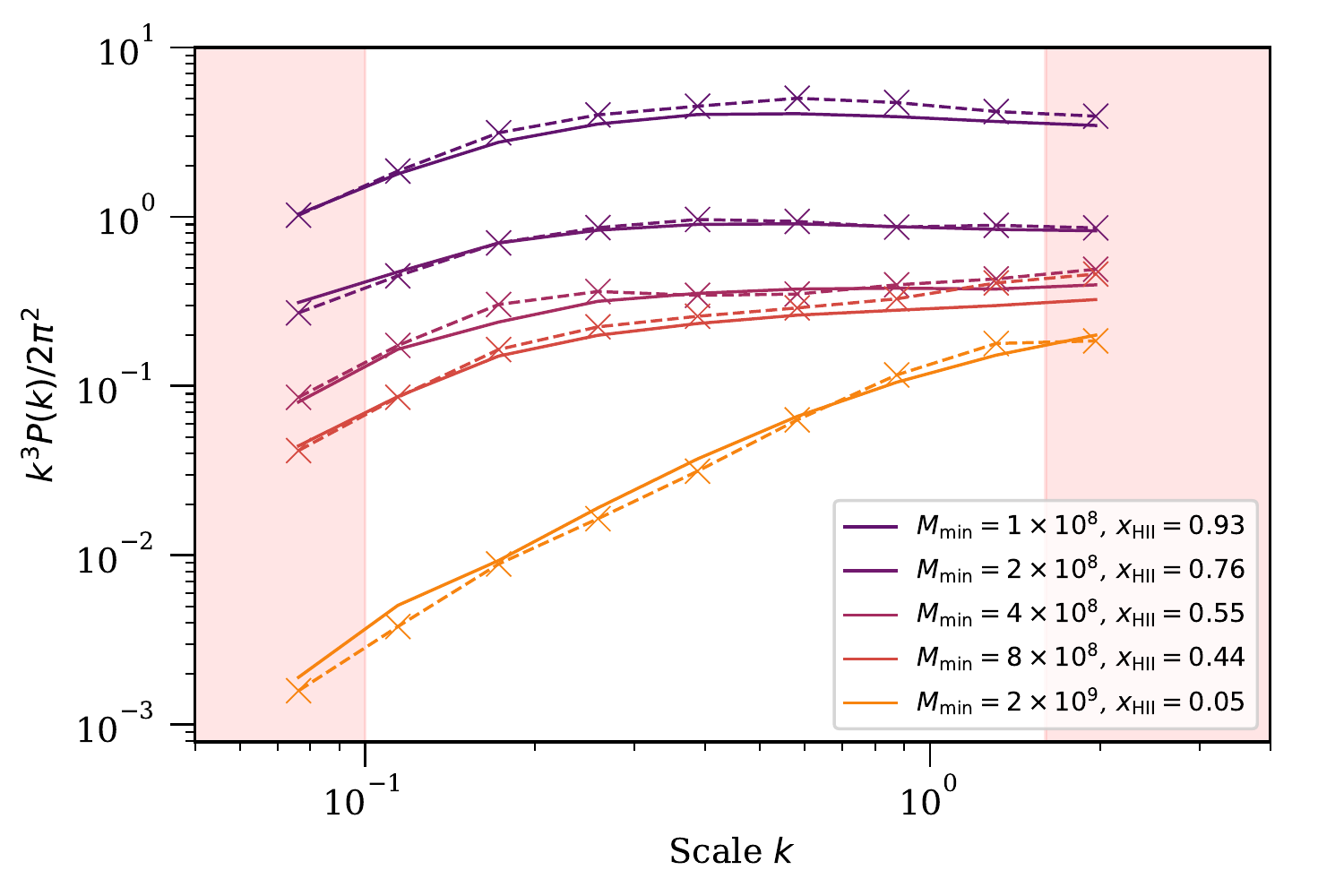}
\caption{\changes{Example emulated and simulated power spectra for a range of $\Mmin$ values at $z=9.5$, for fixed $\Zion = 30$ and $\Rmax = 10$. Solid line shows the simulated power spectra, dotted line shows the predicted power spectra from our best emulator. The ionization fraction for each line is given in the legend.}}
\label{fig:power_spectra_predict_mmin}
\end{figure}

The three-layer multilayer perceptron is the best candidate for emulating \simfast{} behaviour. \tabref{tab:emulator_performance} shows that this emulator makes fast and accurate predictions for the test dataset, taking less than a second to match the true simulation outputs within $4\%$ mean squared error averaged across the whole input parameter space. \figref{fig:best_regions_MLP3} shows that the emulator makes accurate predictions across a wide range of input parameters. Worse performance is seen for MLP emulators using fewer hidden layers: increasing the number of layers allows MLP models to be more flexible, and our results indicate that one- and two-layer MLP models are not flexible enough to fit the simulation outputs as accurately as three-layer models. \changes{Figures \ref{fig:power_spectra_predict_ion-eff} and \ref{fig:power_spectra_predict_mmin} show several example power spectra for a range of $\Zion$ and $\Mmin$ values, also showing the predicted power spectra from this best emulator. The shaded red regions in these figures indicate the ranges of excluded k-values. Given the benefit of most three-layer models over two-layer models, it seems likely that models using four or more layers could provide even closer fit to the training data. We do not investigate such models, given our fixed upper limit on training time. Additionally, the benefit of adding more layers would likely be minimal as there is a clear case of diminishing returns for each additional layer: the best MSE for one layer was $27\%$; two layers gave $4.5\%$ MSE; and three layers gave $3.8\%$. }

The two interpolation models are the worst candidates for emulating \simfast{} behaviour. The nearest-neighbour interpolation model has poor prediction accuracy both in terms of the global MSE value of $290\%$ from \tabref{tab:emulator_performance}, and the local MSE across parameter space shown in \figref{fig:best_regions_NND}. The model uses the nearest-neighbour lookup method of \textit{scipy.spatial.KDTree} which is fast but makes no account of noise or smoothness in the simulation behaviour. The linear interpolation model emulates the \simfast{} behaviour more closely: the global MSE is $17\%$ and the local MSE in \figref{fig:best_regions_LND} shows larger regions of good accuracy. This accuracy is at the expense of much slower prediction times. The nearest neighbour model makes predictions for the whole testing dataset in less than a second, whereas the linear interpolation model takes several hours. Our results indicate that interpolation methods cannot capture the complicated behaviour of \simfast{}, justifying the need for more complex machine learning techniques.

Our sparse Gaussian processes model is a poor emulator candidate. Both the local MSE in \figref{fig:best_regions_SGPR} and the global MSE of $36\%$ are poor. The accuracy of the model would almost certainly be improved by increasing the number of inducing points, which would lessen the matrix inversion approximation. However, training models with $m>2730$ would require more than the allowed CPU time. Our value of $m = 2730$ is chosen as the largest number of inducing points whose model training time does not exceed 156 hours. A hard upper limit of $m <18000$ is found for our 128GB RAM architecture, since values of $m$ larger than this cause a \textit{ResourceError} in \textit{tensorflow}. Moreover, increasing the number of inducing points also increases the prediction time: using $m=910$ takes 16 seconds to make predictions for the testing dataset; using $m=2730$ takes 116 seconds. Increasing the number of inducing points gives better accuracy at the expense of much slower prediction times. 

The support vector machine model is also a poor candidate for a \simfast{} emulator. The global MSE of $32\%$ from \tabref{tab:emulator_performance} is one of the worst. This model also has slow prediction speeds, taking 27 seconds to make predictions of the testing data (100 times slower than the best MLP model). It is possible that using other kernels and doing deeper hyperparameter searches would give better accuracy. Given the long prediction times for these models, we find it unlikely that any SVM models would outperform our best MLP emulator, either in terms of speed or accuracy.

\subsection{Low redshift performance}
\label{sec:low-z-issues}

The prediction accuracy of our emulators is worse for lower redshifts than for higher redshifts. If data for $z<10$ are excluded from performance testing, then all our emulators improve significantly: for instance, the three-layer multilayer perceptron's percentage MSE improves from $3.8\%$ to $1.4\%$. The improved values using only high-redshift power spectra are presented in the third column of \tabref{tab:emulator_performance}. There are two effects that could be causing the worse accuracy at lower redshift, which we discuss here.

First, our emulators differ from those of \mcite{Kern2017} and \mcite{Schmit2018} in that our models are trained using the redshift and $k$-scales as extra input dimensions. Our motivation for including redshift and $k$-scales was to allow for immediate predictions at any redshift or $k$-scale. Without including these input dimensions the trained models would only make prediction at the fixed redshifts and $k$-scales of the training data. Making predictions at other input values with such fixed-input emulators would require further interpolation afterwards. Although using $z$ as an input allows for more flexible predictions, this flexibility is likely a cause of poorer emulation at lower redshifts. Without more computing power or faster training algorithms we would suggest that future attempts to emulate the power spectrum should be done for fixed $z$ inputs.

\begin{figure*}
\centering
\subfloat[Three-layer MLP for $z=8.0$]{
\label{fig:best_regions_MLP3_z8_0}
  \includegraphics[width=0.305\textwidth]{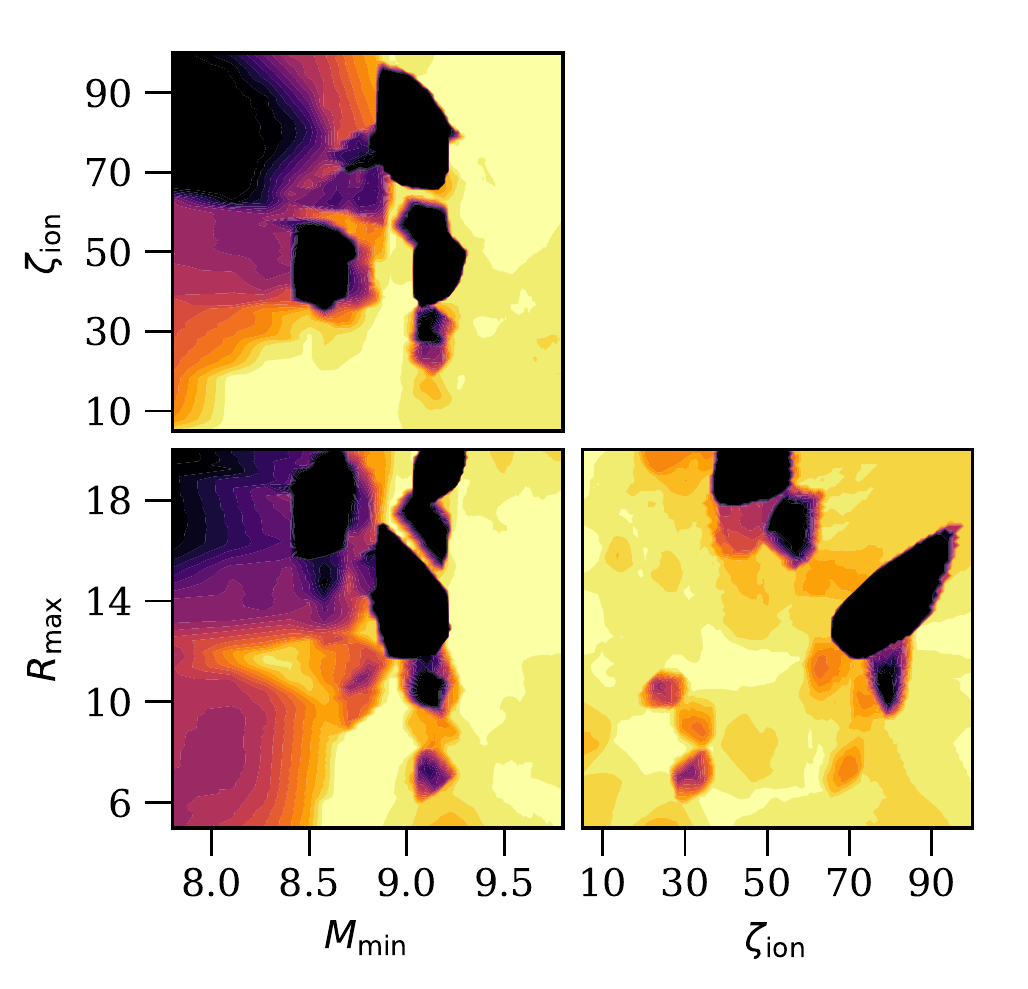}
}
\subfloat[Three-layer MLP for $z=9.5$]{
\label{fig:best_regions_MLP3_z9_5}
  \includegraphics[width=0.305\textwidth]{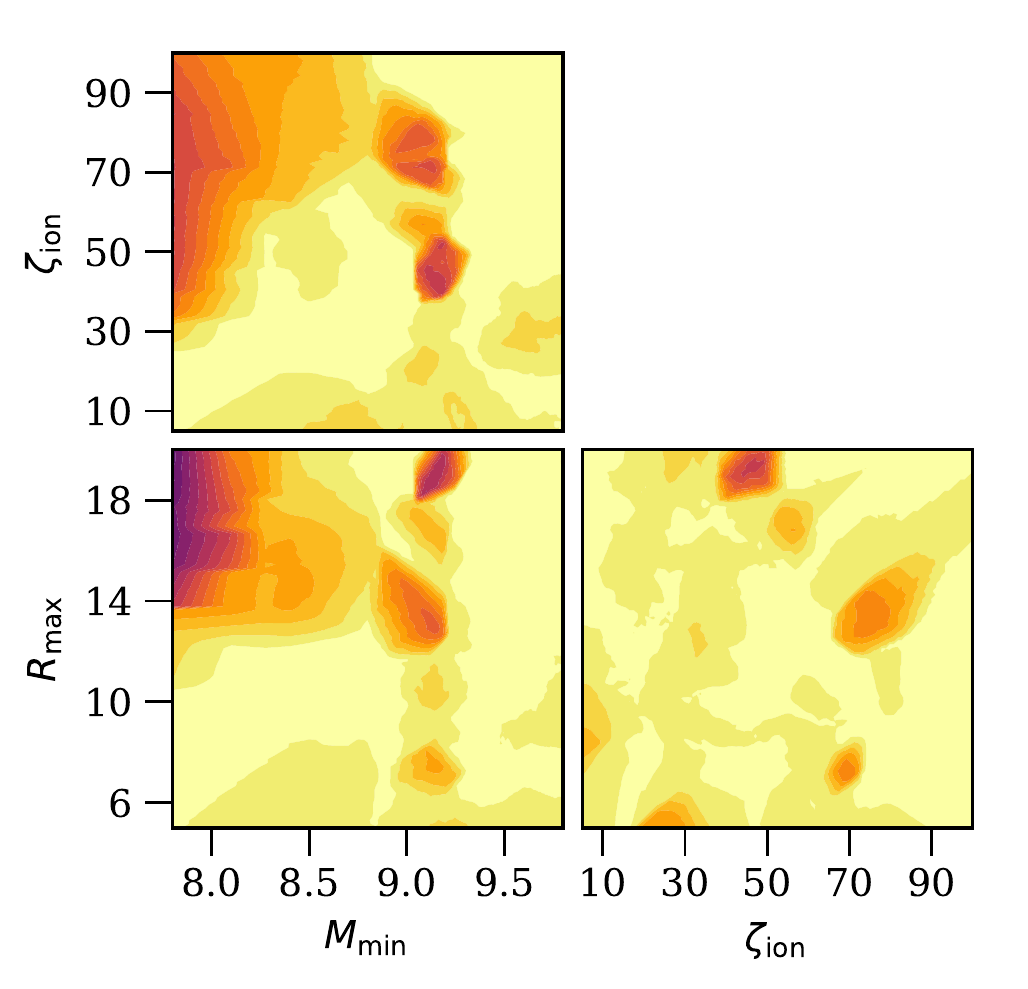}
}
\subfloat[Three-layer MLP for $z=11.0$]{
\label{fig:best_regions_MLP3_z11_0}
  \includegraphics[width=0.39\textwidth]{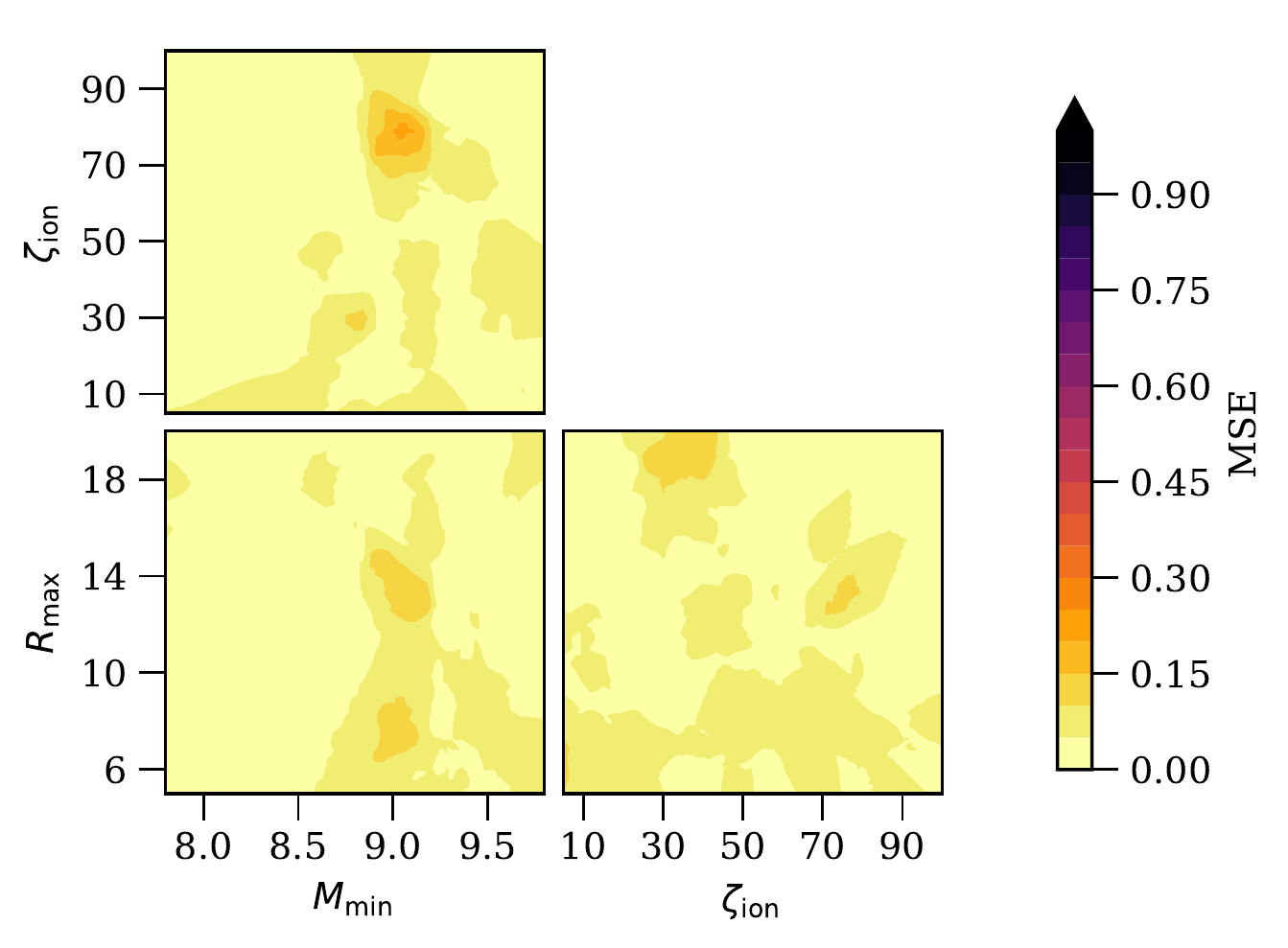}
}
\caption{Mean squared error on testing dataset for the best MLP model as a function of prediction location, similar to \figref{fig:best_regions} but without averaging over redshift. The prediction quality is much worse at low redshift than it is at high redshift, shown by the large darker regions at low redshift. The percentage MSE for $z \geq 11.0$ is better than $5\%$ across almost all of the input parameter space. We omit plotting panels for each $z>11.0$ here as they all look similar to that for $z=11.0$. }
\label{fig:best_region_zs}
\end{figure*}

Secondly, a feature of the actual simulated power spectra could be another source of poor emulator accuracy. The amplitude of the power spectrum in \eqnref{eqn:power_spectrum_definition} is highly sensitive to the global 21cm brightness temperature $\langle \deltatb \rangle$. At low redshifts near the end of reionization, $\langle \deltatb \rangle$ approaches zero (see \mcitealt{PritchardLoeb2012} for a review). This low value of $\langle \deltatb \rangle$ amplifies even small fluctuations in $\deltatb$, causing a sharp increase in the fluctuation field $\Delta T_b(\vec{r})$ from which the power spectrum is calculated. Soon after, reionization finishes and $\Delta T_b(\vec{r})=0$ everywhere so that the amplitude of $P_{\Deltatb}(k)$ jumps suddenly from high-amplitude to zero-amplitude. These two sudden features are difficult to emulate: the sharp increase in power spectrum amplitude near the end of reionization, and the sudden drop thereafter from high-amplitude to zero-amplitude.

\begin{figure}
\includegraphics[width=\columnwidth]{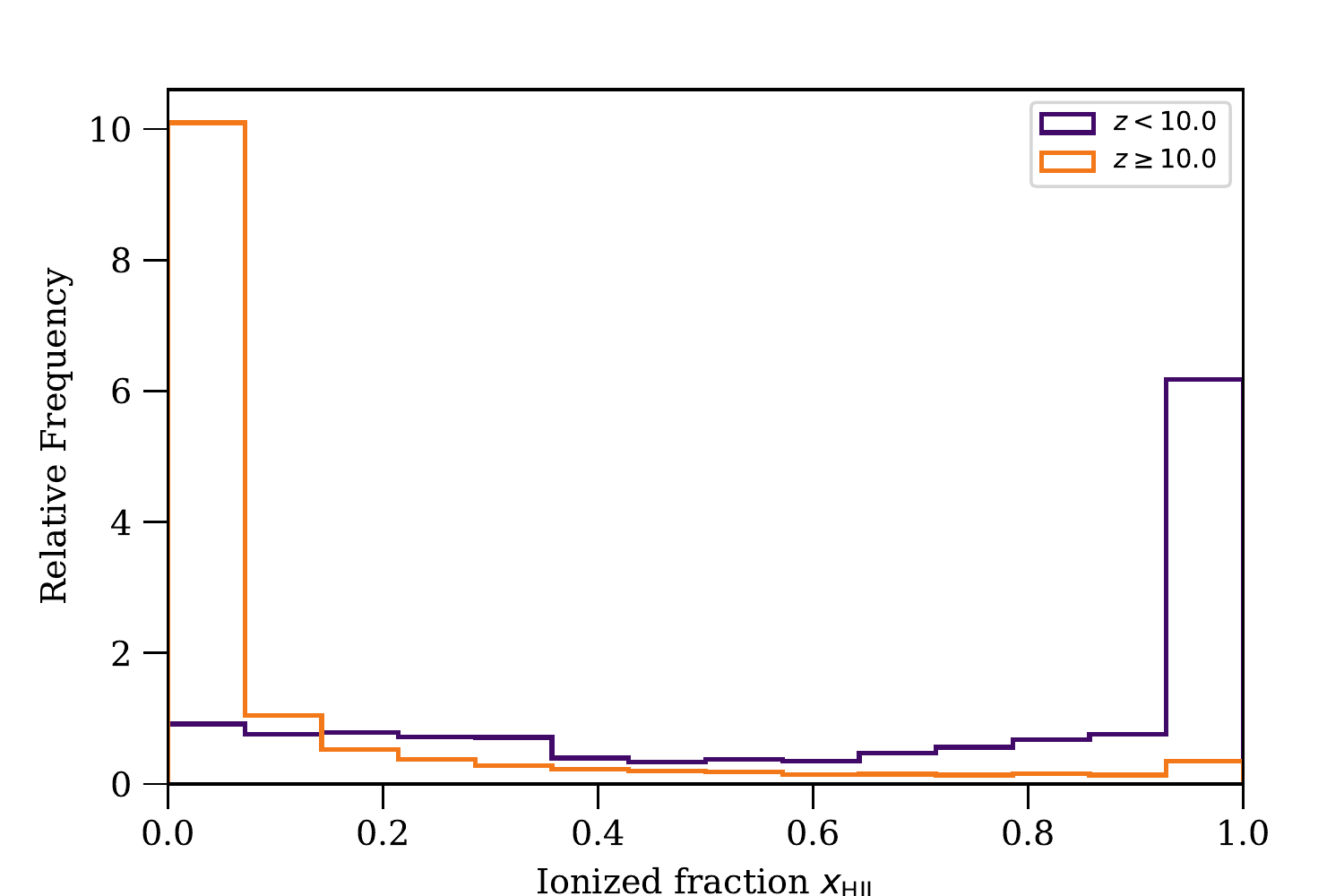}
\caption{\changes{Normalised histogram of the ionization fraction for all simulations, for low redshifts ($z<10$) and for higher redshifts ($z \geq 10$). The 21cm power spectrum is sensitive to the neutral fraction. The fact that many simulations are fully ionized for $z<10$ could be one reason for the poor performance of our emulators at low redshifts.}}
\label{fig:xhii_hist}
\end{figure}

\changes{We investigate the low redshift behaviour in two ways. First, we plot the local prediction performance of our best emulator at each redshift in the training data. \figref{fig:best_region_zs} shows the best emulator's local MSE separately for the three lowest redshifts $z=\{8.0,\ 9.5,\ 11.0\}$. For $z \geq 10$, very few of the simulations in our training data have completed reionization and thus very few contain the associated problematic zero-valued power spectra. In \figref{fig:best_regions_MLP3_z11_0} and for higher redshifts, the local performance is consistently good across the parameter space. At $z=9.5$ some simulations are near the end of reionization, and the local performance in \figref{fig:best_regions_MLP3_z9_5} begins to show regions of poorer predictions. By $z=8$, a large number of simulations have completed reionization. These regions have zero-amplitude power spectra and show much worse performance, in particular those with high $\Zion$ and low $\Mmin$ in \figref{fig:best_regions_MLP3_z8_0}.}

\changes{Our second method to examine the low redshift behaviour is to observe how many simulations have finished reionization at each redshift. \figref{fig:xhii_hist} shows normalised histograms of the global ionization fraction values for all simulations in our data, separating into lower redshifts ($z<10$) and higher redshifts ($z \geq 10$). For higher redshifts, most simulations are near the start of reionization, with ionization fractions  $x_{\mathrm{HII}} < 0.4$. For lower redshifts, many models have finished reionization with ionization fractions nearing $x_{\mathrm{HII}} = 1.0$. This difference causes the low redshift power spectra to contain sudden features which are difficult to emulate.} Training using $\deltatb - \langle \deltatb \rangle$ as the target values rather than $\Deltatb$ could remove the sudden changes in power spectra magnitudes and would be easier to emulate. We note for instance that \mcite{Kern2017} were able to emulate down to $z=5$ without reporting any issues for emulating these redshifts.

\section{Use case: mapping between \simfast{} and \cmfast{}}
\label{sec:bias}

\cmfast{} and \simfast{} are two common semi-numerical simulations for generating predicted 21cm maps during the Epoch of Reionization. In this section we use our best emulator to investigate the extent to which the two simulations give similar outputs for similar inputs. Our motivation for this is to demonstrate a method for creating a mapping between any two simulations which generate the same output statistic. In particular, this method could be extended to give a mapping between \simfast{} power spectra and those from more accurate (but slower) three-dimensional radiative transfer simulations, such as \cray{} (\mcitealt{c2ray}). \changes{Although numerical simulations have different input parameters to semi-numerical simulations, it would still be possible to map between them by finding the parameters which best match their output power spectra.} When analysing huge datasets, \simfast{} could be used to give coarse constraints on reionization parameters. Using the mapping between \simfast{} and the more accurate numerical simulation, the coarse contours could be mapped to their equivalent regions of the numerical simulation inputs. This would allow more detailed exploration of this smaller region of parameter space with the numerical simulations.

\subsection{\cmfast{}}
Here we describe the default procedure of \cmfast{}, in particular highlighting how it differs from the \simfast{} algorithms described earlier in \secref{sec:simfast21}. In the following section we discuss which of these differences we retain when creating the mapping. The linear and non-linear density fields in \cmfast{} are seeded in the same way as in \simfast{}. 

The first difference between the simulations is the method for calculating collapse fractions from the non-linear density field. \cmfast{} does not resolve individual halos, but rather calculates the collapse fraction directly from the non-linear density field following the model of spherical collapse from \mcite{PressSchechter1974}. In order to match the more accurate ellipsoidal collapse model from \mcite{Sheth2001}, \cmfast{} afterwards normalises the spherical collapse fractions so that their average value matches that expected from ellipsoidal collapse.

The second difference is the method for calculating the ionization fraction. Both simulations calculate the ionization fraction by determining whether the collapsed matter in a region emits enough photons to ionize the surrounding matter. If there are enough photons, then \simfast{} paints the entire spherical region as ionized using the fully overlapping-spheres method in \mcite{Mesinger2007}, whereas the default \cmfast{} algorithm is to paint only the central pixel of the region \mcite{Zahn2007}. The latter method is much faster but the algorithms give a considerably different reionization history for the same inputs (see \mcitealt{hutter2018}). \cmfast{} has an option to match the method of \simfast{}. 

The final difference is in the evolution of the parameter $\Mmin$. The default \cmfast{} implementation allows the minimum halo mass $\Mmin$ to evolve with redshift by setting a minimum virial temperature $\Tvir$ for ionizing photons. 

\subsection{Matching reionization histories}

Using identical reionization and cosmological parameters, and keeping all other input parameters at their default values from the \textit{GitHub} packages, \cmfast{} version 1.2\footnote{https://github.com/andreimesinger/21cmFAST} and \simfast{} version 1.0 result in different reionization histories, as expected due to the different default bubble-finding algorithms. Our motivation in this section is to demonstrate a method for mapping between the input parameters of two similar (but not identical) simulations. Using the default implementations, the output power spectra of the two simulations at a single fixed redshift are not comparable because the two simulations have reached different stages of reionization. Before making the mapping, we chose input parameters of \cmfast{} which more closely matched the \simfast{} algorithm, so that the output power spectra are similar enough that making a mapping is meaningful, but not so similar that they give identical results. The following is a list of significant input parameters in \cmfast{} that we adjusted from the default values:

\begin{enumerate}
\item $\text{FIND\_BUBBLE\_ALGORITHM} = 1$
\item $\text{ION\_Tvir\_MIN} = -1$, instead using ION\_M\_MIN
\item $\text{INHOMO\_RECO} = 0$
\end{enumerate}

\begin{figure}
\includegraphics[width=\columnwidth]{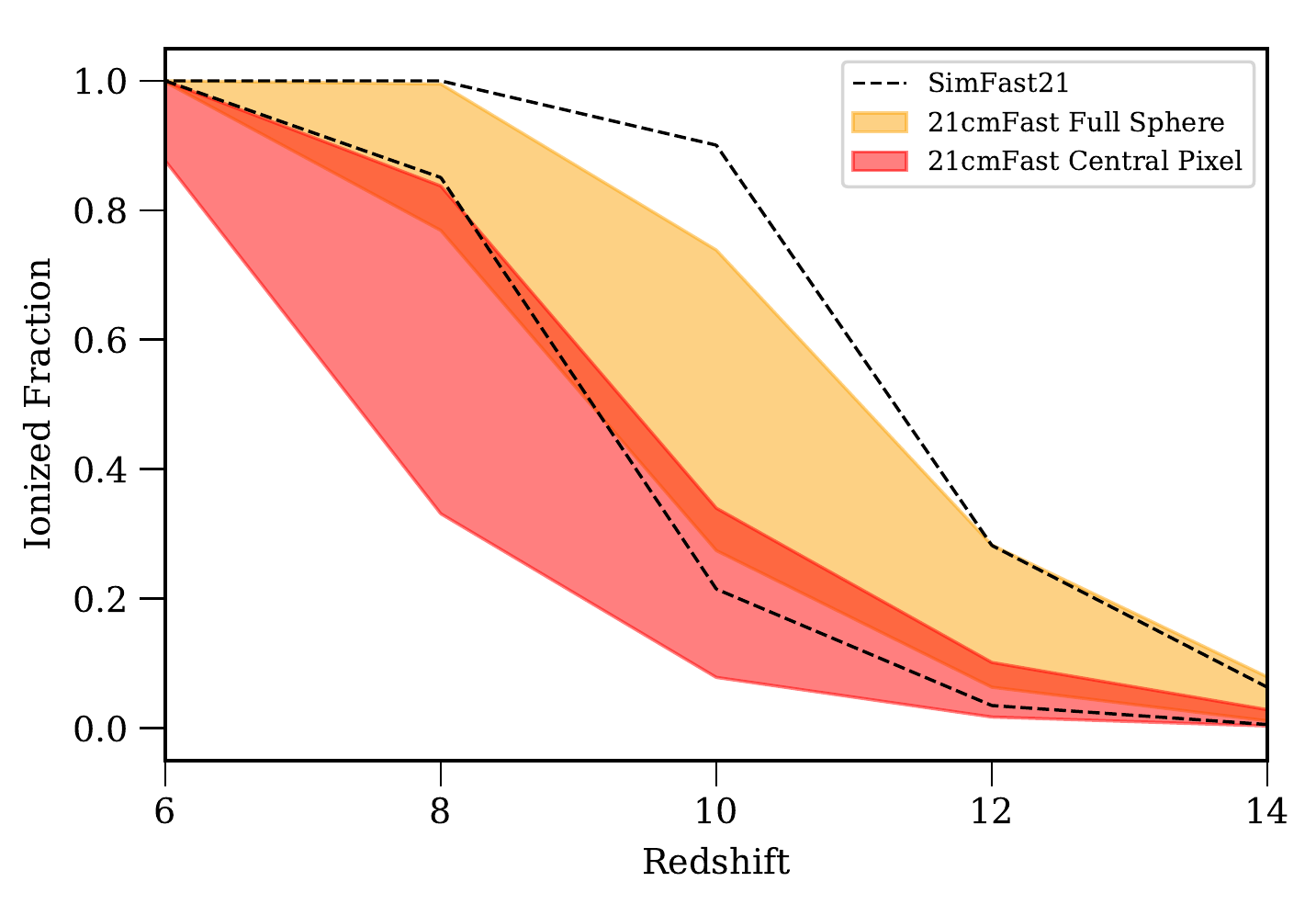}
\caption{Ranges of reionization histories that result from \simfast{} and \cmfast{}, with $\Mmin$ varying from $10^8 M_{\odot}$ to $10^9 M_{\odot}$. The region between the black dotted curves indicates the range of histories from \simfast{}. The two coloured regions show the range of histories from \cmfast{}, both before (darker red) and after (lighter orange) matching the algorithms. The other reionization parameters are fixed at $\Zion = 30.0$ and $\Rmax=10.0$. The bubble-finding algorithm has a significant impact on the resulting reionization history, and even after matching algorithms there is a slight difference between \simfast{} and \cmfast{}.}
\label{fig:reion_histories}
\end{figure}

\noindent Appendix \ref{appendix:parameters} lists all parameters used in both simulations for repeatability. The most significant change from default was in the algorithm for finding ionized bubbles, setting $\text{FIND\_BUBBLE\_ALGORITHM} = 1$. Without making a judgement on which method is more realistic we used the \simfast{} algorithm of painting the entire sphere as ionized, rather than painting only the central pixel. We fix the minimum mass $\Mmin$ for collapse using ION\_M\_MIN, rather than using the default \cmfast{} functionality of a fixed virial temperature $\Tvir$ using ION\_Tvir\_MIN. We also turn off calculations involving inhomogeneous recombinations by setting $\text{INHOMO\_RECO} = 0$, since the version of \simfast{} that we use does not have this option (although later versions do, see \citealt{Hassan2016}).  \figref{fig:reion_histories} shows the resulting ranges of reionization histories from a spread of minimum halo mass scenarios between $10^{8} M_{\odot} - 10^{9} M_{\odot}$. Each minimum mass scenario is averaged across five realisations. The histories are shown for \simfast{} (dotted) and for \cmfast{} with both bubble-finding algorithms: ionising the central pixel only (darker red region) and ionising the full sphere (lighter orange region) to match \simfast{}. The only remaining major differences between the default \simfast{} simulation and the changed \cmfast{} simulation are in the specifics of implementation discussed above. \figref{fig:reion_histories} shows that the differences in implementation still result in different reionisation histories even after matching the bubble-finding algorithms, although the bubble-finding algorithm is the most dominant effect.

\subsection{Determining a mapping between simulations}

Here we describe how we use our best emulator to determine a mapping between the inputs of our modified \cmfast{} and the inputs of \simfast{}. We use the same $k$-space restrictions as in \secref{sec:k-range-restrictions}, using only $0.1 \leq k \leq 2.0$ since the large scales are subject to foregrounds and the small scales are subject to shot noise from the finite simulation resolution. We also restrict our comparisons to higher redshifts $z \geq 10$ for which our emulator exhibits higher prediction accuracy. We emphasise that this is a proof-of-concept method showing how to make a mapping between simulations solely using the output power spectra.

\begin{figure}
\includegraphics[width=\columnwidth]{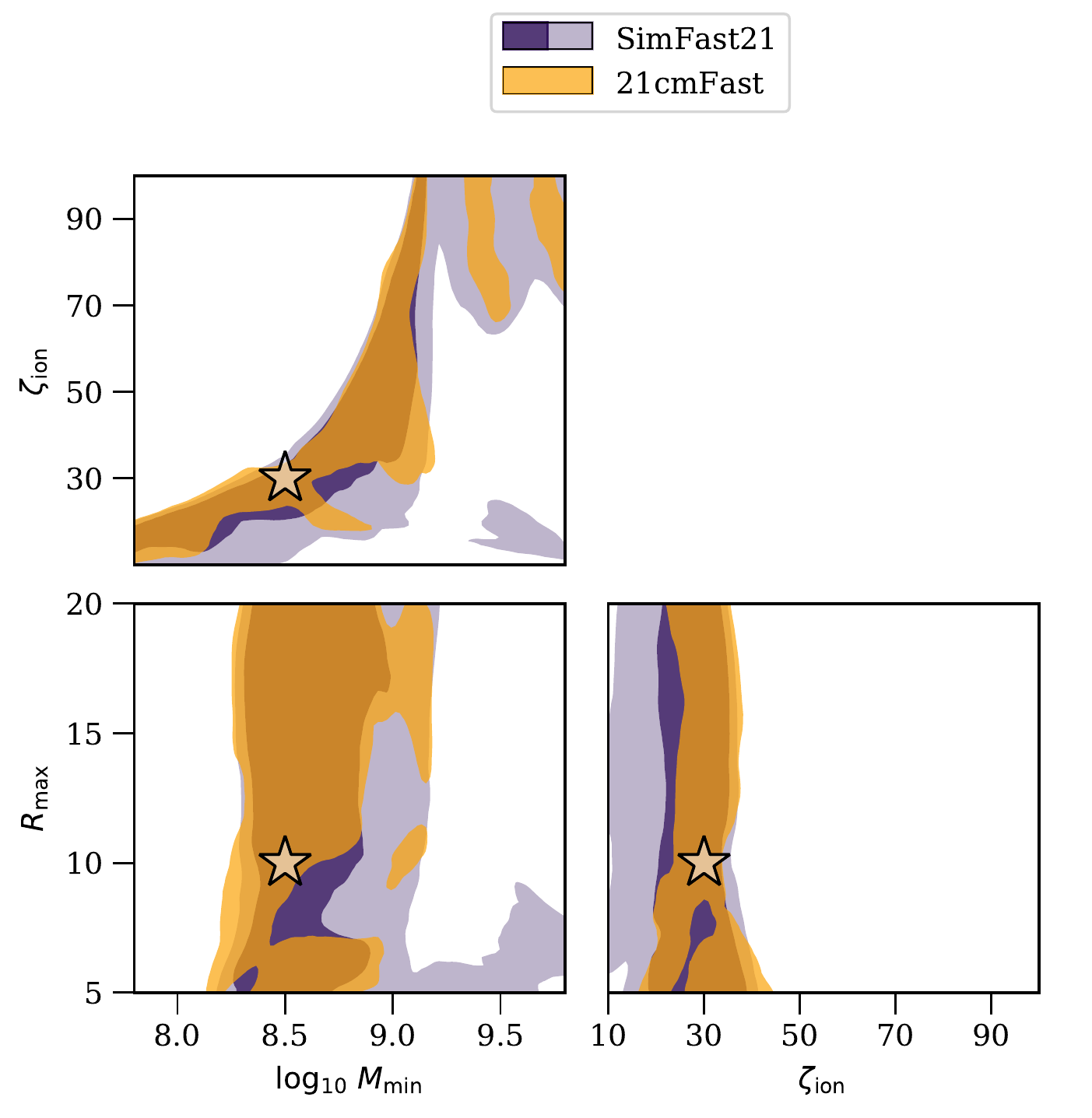}
\caption{Mean squared error between emulated \simfast{} power spectra and measured power spectra from both simulations. The star indicates the fixed simulation parameters. The orange contour indicates the regions where emulated \simfast{} power spectra are within $30\%$ MSE of the fully-simulated \cmfast{} power spectra. For comparison, the purple contours indicate the same regions for comparing emulated \simfast{} power spectra with fully-simulated \simfast{} power spectra, using $30\%$ MSE (lighter contour) and $15\%$ MSE (darker contour).}
\label{fig:similarity_3d}
\end{figure}

\changes{\figref{fig:similarity_3d} shows an example of one such mapping. We explain how to interpret the mapping here. Suppose a reference \cmfast{} simulation has already been run using the parameters specified by the white star: namely, a \cmfast{} simulation with parameters $\Mmin=3 \times 10^8 M_{\odot}$, $\Zion = 30.0$, $\Rmax = 10 \text{Mpc}$. According to the mapping in \figref{fig:similarity_3d}, any \simfast{} simulation using parameters within the orange contour will result in power spectra which are similar to the reference \cmfast{} spectra. We classify two simulations as similar if the mean-squared error between their output power spectra is lower than $30\%$. The orange contour thus shows the region of \simfast{} parameters which should be used, if the desired result is to exhibit similar power spectra to the reference \cmfast{} simulation. We generate the reference power spectra in \figref{fig:similarity_3d} by running five \cmfast{} simulations, and taking the average to reduce the effect of sample variance. We then generate the contours by using our emulator to run a large number of simulations across the whole parameter space. For each emulated simulation, we calculate the mean-squared error of its power spectra compared to the reference simulation. Note that our emulator makes these predictions in seconds, rather than the several months that would be needed to run the same number of full simulations. We refer to this type of figure as a similarity plot. Most importantly, if the orange contour does not overlap with the white star, then this indicates that \simfast{} and \cmfast{} result in significantly different output power spectra for the same input parameters.} 

\changes{Two features of the orange contours are immediately apparent. First, the extended contours in the $\Rmax$ direction. The $\Rmax$ parameter is known to have little effect on the output power spectra for our high redshifts \mcitep{21cmFast}. This is an inherent property of the power spectrum, regardless of which simulation is used. A second clear feature is the large curved contour in the $\Mmin$-$\Zion$ parameter space. We investigated both features, to confirm whether they arise as an inherent property of the power spectrum itself, or if they arise from differences in the two simulations. To do this, we perform the same similarity analysis as above, but using \simfast{} itself as the reference simulation. The purple contours in \figref{fig:similarity_3d} then give the regions of \simfast{} parameters which result in similar power spectra to the reference \simfast{} simulation.  The lighter purple contours use a MSE threshold of $30\%$. The darker purple contours use a stricter threshold of $15\%$ MSE. The curved feature appears in both orange and purple contours, indicating that it is not due to a difference in the simulations. This curved degeneracy has been observed previously, see for example \mcite{21CMMC2015} and \mcite{Schmit2018}. Note that we do not include a dark orange contour for the the stricter $15\%$ MSE threshold because the power spectra for \cmfast{} differ from those of \simfast{} enough that no \cmfast{} contours are visible for an MSE threshold of $15\%$. }

\begin{figure*}
\includegraphics[width=\textwidth]{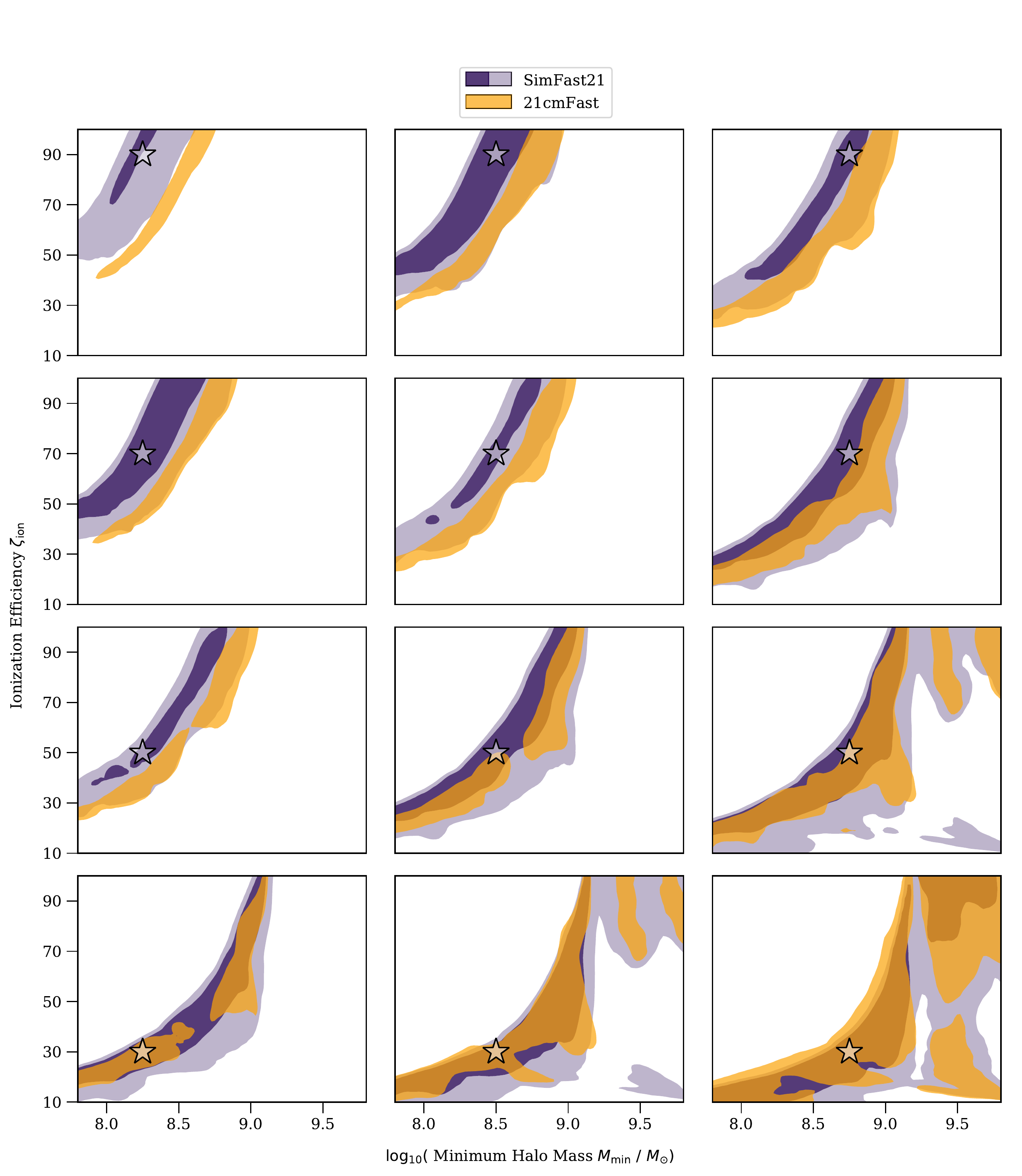}
\caption{Similarity plots between emulated \simfast{} power spectra and fully-simulated power spectra from \cmfast{} (orange contours) and \simfast{} (purple contours). In each panel, the white star indicates the scenario parameters of the fully-simulated power spectra. The orange contour shows the regions in which emulated \simfast{} power spectra differ by less than $30 \%$ from the fully-simulated \cmfast{} power spectra. The lighter- and darker-purple contours show the equivalent regions for comparing emulated \simfast{} power spectra to fully-simulated \simfast{} power spectra, within $30 \%$ and $15 \%$ MSE respectively. An offset can be seen for several of these different scenarios.}
\label{fig:compare_2d_3x4}
\end{figure*}

\changes{\figref{fig:compare_2d_3x4} shows similarity plots for several other reference simulations, where the parameters for each reference simulation is again indicated by the location of the white star. We show the contours in the two-dimensional $\Mmin$-$\Zion$ space, ignoring the less interesting $\Rmax$ direction. We find that the orange contour does not always lie on top of the white star. This indicates that \simfast{} and \cmfast{} do not always result in similar output power spectra. We use the same contour levels as in \figref{fig:similarity_3d}, namely $15\%$ and $30\%$ for the darker and lighter purple contours, and $30\%$ for the \cmfast{} contours. Again, no $15\%$ MSE contour is shown for \cmfast{} because the \simfast{} power spectra differ from the \cmfast{} power spectra by more than $15\%$ everywhere.}

\changes{For several of these scenarios, there is an offset between the orange contour and the white star. The offset is small near the canonical parameters in the central panels, but gets larger at lower $\Mmin$ and higher $\Zion$.} The most likely reason for this offset is the difference in the reionization histories. This offset would mean that the choice of using \simfast{} or using \cmfast{} would affect the outcome of parameter estimation methods, such as maximising $\chi^2$ values in \mcite{Shimabukuro2017}, or using MCMC methods as in \mcite{Schmit2018} and \mcite{Kern2017}. \changes{Note that the two simulations in this comparison needn't share the same types of input parameters. For instance, it would be possible to generate the reference power spectra using a numerical radiative transfer simulation, and determine how its inputs map to \simfast{}.}

\section{Conclusions}
\label{sec:conclusions}

Fast modelling of the 21cm signal will become a significant problem in analysing the huge datasets from upcoming radio interferometry experiments. Ideally, we would be able to compare numerical radiative transfer simulations with these data. Current numerical simulations are too slow to sample the input parameter space efficiently. Semi-numerical simulations are faster but can still only be used to constrain a small number of parameters. One potential solution to this problem is to replace current semi-numerical simulations with emulated models, reproducing the simulation outputs in a fraction of the original simulation time.

In this paper, we train and compare emulators using five different machine learning techniques. The two naive interpolation methods are not feasible as emulators, since they have either slow prediction times (linear interpolation model) or poor accuracy (nearest neighbour interpolation model). Of the three more sophisticated models, one model performs much better than the others: the multilayer perceptron. This trained model makes predictions of the outputs from 500 \simfast{} simulations to within 4\% mean squared error averaged across all output points, reducing the modelling time from around 3000 hours to less than a second. If CPU training time is not a factor, then the accuracy of the sparse Gaussian processes regression or support vector machine models could potentially be improved with deeper hyperparameter searches. However, given their already relatively long prediction times and the accurate performance of the multilayer perceptron, these models are unlikely to give an improvement over the three-layer multilayer perceptron.

Our emulators use redshift and $k$-scales as extra input dimensions. This makes the models more flexible but gives rise to less accurate emulation especially near the end of reionization at lower redshifts. We also use $\Deltatb = \frac{\deltatb}{\langle \deltatb \rangle} - 1$ as the target values of our emulators. This gives rise to sudden features in the power spectra near at the end of reionization and is harder to emulate than using $\deltatb - \langle \deltatb \rangle$.

We use our best emulator to determine a relationship between two different reionization algorithms, using \simfast{} and a version of \cmfast{} with non-default inputs. We find some noticeable offsets in which input parameters match the power spectra outputs of \simfast{} with those of \cmfast{}. We provide a graphical description of how this offset depends on location in parameter space, so that users could roughly determine which \simfast{} input parameters should be used if the desired result is to match the 21cm power spectrum of an existing \cmfast{} simulation. Although our results are for a version of \cmfast{} with non-default inputs, this method has potential for bridging between fast semi-numerical simulations and more accurate three-dimensional radiative transfer code such as \cray{} (\mcitealt{c2ray}). However, \mcite{Suman2014} noted that there can be a 25\% difference between the power spectrum outputs of \cray{} and semi-numerical codes. Given this discrepancy, it is likely that mapping between numerical and semi-numerical simulations will be considerably more challenging and it may be necessary to emulate numerical codes directly.

\section*{Acknowledgements}
Many thanks to Mario Santos and Andrei Mesinger for their helpful comments on a draft version of this paper. WDJ is supported by the Science and Technology Facilities Council (ST/M503873/1) and from the European Community through the DEDALE grant (contract no. 665044) within the H2020 Framework Program of the European Commission. CAW acknowledges financial support from the European Research Council under ERC grant number 638743-FIRSTDAWN (held by Jonathan Pritchard). FBA acknowledges support from the DEDALE grant, from the UK Science and Technology Research Council (STFC) grant ST/M001334/1, and from STFC grant ST/P003532/1.

\bibliography{emulate}

\begin{thebibliography}{}
\makeatletter
\relax
\def\mn@urlcharsother{\let\do\@makeother \do\$\do\&\do\#\do\^\do\_\do\%\do\~}
\def\mn@doi{\begingroup\mn@urlcharsother \@ifnextchar [ {\mn@doi@}
  {\mn@doi@[]}}
\def\mn@doi@[#1]#2{\def\@tempa{#1}\ifx\@tempa\@empty \href
  {http://dx.doi.org/#2} {doi:#2}\else \href {http://dx.doi.org/#2} {#1}\fi
  \endgroup}
\def\mn@eprint#1#2{\mn@eprint@#1:#2::\@nil}
\def\mn@eprint@arXiv#1{\href {http://arxiv.org/abs/#1} {{\tt arXiv:#1}}}
\def\mn@eprint@dblp#1{\href {http://dblp.uni-trier.de/rec/bibtex/#1.xml}
  {dblp:#1}}
\def\mn@eprint@#1:#2:#3:#4\@nil{\def\@tempa {#1}\def\@tempb {#2}\def\@tempc
  {#3}\ifx \@tempc \@empty \let \@tempc \@tempb \let \@tempb \@tempa \fi \ifx
  \@tempb \@empty \def\@tempb {arXiv}\fi \@ifundefined
  {mn@eprint@\@tempb}{\@tempb:\@tempc}{\expandafter \expandafter \csname
  mn@eprint@\@tempb\endcsname \expandafter{\@tempc}}}

\bibitem[\protect\citeauthoryear{Abadi et~al.,}{Abadi
  et~al.}{2015}]{tensorflow2015-whitepaper}
Abadi M.,  et~al., 2015, {TensorFlow}: Large-Scale Machine Learning on
  Heterogeneous Systems, \url {https://www.tensorflow.org/}

\bibitem[\protect\citeauthoryear{{Ali} et~al.,}{{Ali} et~al.}{2015}]{PAPER}
{Ali} Z.~S.,  et~al., 2015, \mn@doi [\apj] {10.1088/0004-637X/809/1/61}, \href
  {http://adsabs.harvard.edu/abs/2015ApJ...809...61A} {809, 61}

\bibitem[\protect\citeauthoryear{{Alvarez} \& {Abel}}{{Alvarez} \&
  {Abel}}{2012}]{Alvarez&Abel2012}
{Alvarez} M.~A.,  {Abel} T.,  2012, \mn@doi [\apj]
  {10.1088/0004-637X/747/2/126}, \href
  {http://adsabs.harvard.edu/abs/2012ApJ...747..126A} {747, 126}

\bibitem[\protect\citeauthoryear{Barber, Dobkin  \& Huhdanpaa}{Barber
  et~al.}{1996}]{qhull}
Barber C.~B.,  Dobkin D.~P.,   Huhdanpaa H.,  1996, ACM TRANSACTIONS ON
  MATHEMATICAL SOFTWARE, 22, 469

\bibitem[\protect\citeauthoryear{{Barkana} \& {Loeb}}{{Barkana} \&
  {Loeb}}{2001}]{Barkana2001}
{Barkana} R.,  {Loeb} A.,  2001, \mn@doi [\physrep]
  {10.1016/S0370-1573(01)00019-9}, \href
  {http://adsabs.harvard.edu/abs/2001PhR...349..125B} {349, 125}

\bibitem[\protect\citeauthoryear{{Datta}, {Bowman}  \& {Carilli}}{{Datta}
  et~al.}{2010}]{Datta2010}
{Datta} A.,  {Bowman} J.~D.,   {Carilli} C.~L.,  2010, \mn@doi [\apj]
  {10.1088/0004-637X/724/1/526}, \href
  {http://adsabs.harvard.edu/abs/2010ApJ...724..526D} {724, 526}

\bibitem[\protect\citeauthoryear{{DeBoer} et~al.,}{{DeBoer}
  et~al.}{2017}]{HERA}
{DeBoer} D.~R.,  et~al., 2017, \mn@doi [\pasp]
  {10.1088/1538-3873/129/974/045001}, \href
  {http://adsabs.harvard.edu/abs/2017PASP..129d5001D} {129, 045001}

\bibitem[\protect\citeauthoryear{Furlanetto, Zaldarriaga  \&
  Hernquist}{Furlanetto et~al.}{2004}]{Furlanetto2004b}
Furlanetto S.~R.,  Zaldarriaga M.,   Hernquist L.,  2004, The Astrophysical
  Journal, 613, 1

\bibitem[\protect\citeauthoryear{{Furlanetto}, {Oh}  \& {Briggs}}{{Furlanetto}
  et~al.}{2006}]{Furlanetto2006}
{Furlanetto} S.~R.,  {Oh} S.~P.,   {Briggs} F.~H.,  2006, \mn@doi [\physrep]
  {10.1016/j.physrep.2006.08.002}, \href
  {http://adsabs.harvard.edu/abs/2006PhR...433..181F} {433, 181}

\bibitem[\protect\citeauthoryear{{Gillet}, {Mesinger}, {Greig}, {Liu}  \&
  {Ucci}}{{Gillet} et~al.}{2018}]{Gillet2018}
{Gillet} N.,  {Mesinger} A.,  {Greig} B.,  {Liu} A.,   {Ucci} G.,  2018,
  preprint, \href {http://adsabs.harvard.edu/abs/2018arXiv180502699G} {}
  (\mn@eprint {arXiv} {1805.02699})

\bibitem[\protect\citeauthoryear{{Greig} \& {Mesinger}}{{Greig} \&
  {Mesinger}}{2015}]{21CMMC2015}
{Greig} B.,  {Mesinger} A.,  2015, \mn@doi [\mnras] {10.1093/mnras/stv571},
  \href {http://adsabs.harvard.edu/abs/2015MNRAS.449.4246G} {449, 4246}

\bibitem[\protect\citeauthoryear{{Greig} \& {Mesinger}}{{Greig} \&
  {Mesinger}}{2018}]{Greig2018}
{Greig} B.,  {Mesinger} A.,  2018, in {Jeli{\'c}} V.,  {van der Hulst} T.,
  eds,  IAU Symposium Vol. 333, Peering towards Cosmic Dawn. pp 18--21
  (\mn@eprint {arXiv} {1705.03471}), \mn@doi{10.1017/S1743921317011103}

\bibitem[\protect\citeauthoryear{{Greig}, {Mesinger}  \& {Pober}}{{Greig}
  et~al.}{2016}]{Greig2016}
{Greig} B.,  {Mesinger} A.,   {Pober} J.~C.,  2016, \mn@doi [\mnras]
  {10.1093/mnras/stv2618}, \href
  {http://adsabs.harvard.edu/abs/2016MNRAS.455.4295G} {455, 4295}

\bibitem[\protect\citeauthoryear{{Hassan}, {Dav{\'e}}, {Finlator}  \&
  {Santos}}{{Hassan} et~al.}{2016}]{Hassan2016}
{Hassan} S.,  {Dav{\'e}} R.,  {Finlator} K.,   {Santos} M.~G.,  2016, \mn@doi
  [\mnras] {10.1093/mnras/stv3001}, \href
  {http://adsabs.harvard.edu/abs/2016MNRAS.457.1550H} {457, 1550}

\bibitem[\protect\citeauthoryear{{Hassan}, {Dav{\'e}}, {Finlator}  \&
  {Santos}}{{Hassan} et~al.}{2017}]{Hassan2017}
{Hassan} S.,  {Dav{\'e}} R.,  {Finlator} K.,   {Santos} M.~G.,  2017, \mn@doi
  [\mnras] {10.1093/mnras/stx420}, \href
  {http://adsabs.harvard.edu/abs/2017MNRAS.468..122H} {468, 122}

\bibitem[\protect\citeauthoryear{{Hutter}}{{Hutter}}{2018}]{hutter2018}
{Hutter} A.,  2018, \mn@doi [\mnras] {10.1093/mnras/sty683}, \href
  {http://adsabs.harvard.edu/abs/2018MNRAS.477.1549H} {477, 1549}

\bibitem[\protect\citeauthoryear{Jones, Oliphant, Peterson  et~al.}{Jones
  et~al.}{2001}]{scipy}
Jones E.,  Oliphant T.,  Peterson P.,   et~al., 2001, {SciPy}: Open source
  scientific tools for {Python}, \url {http://www.scipy.org/}

\bibitem[\protect\citeauthoryear{{Kern}, {Liu}, {Parsons}, {Mesinger}  \&
  {Greig}}{{Kern} et~al.}{2017}]{Kern2017}
{Kern} N.~S.,  {Liu} A.,  {Parsons} A.~R.,  {Mesinger} A.,   {Greig} B.,  2017,
  \mn@doi [\apj] {10.3847/1538-4357/aa8bb4}, \href
  {http://adsabs.harvard.edu/abs/2017ApJ...848...23K} {848, 23}

\bibitem[\protect\citeauthoryear{{Kingma} \& {Ba}}{{Kingma} \&
  {Ba}}{2014}]{Adam}
{Kingma} D.~P.,  {Ba} J.,  2014, preprint, \href
  {http://adsabs.harvard.edu/abs/2014arXiv1412.6980K} {} (\mn@eprint {arXiv}
  {1412.6980})

\bibitem[\protect\citeauthoryear{{Kulkarni}, {Choudhury}, {Puchwein}  \&
  {Haehnelt}}{{Kulkarni} et~al.}{2016}]{Girish2016}
{Kulkarni} G.,  {Choudhury} T.~R.,  {Puchwein} E.,   {Haehnelt} M.~G.,  2016,
  \mn@doi [\mnras] {10.1093/mnras/stw2168}, \href
  {http://adsabs.harvard.edu/abs/2016MNRAS.463.2583K} {463, 2583}

\bibitem[\protect\citeauthoryear{{Liu}, {Pritchard}, {Allison}, {Parsons},
  {Seljak}  \& {Sherwin}}{{Liu} et~al.}{2016}]{Liu2016}
{Liu} A.,  {Pritchard} J.~R.,  {Allison} R.,  {Parsons} A.~R.,  {Seljak} U.,
  {Sherwin} B.~D.,  2016, \mn@doi [\prd] {10.1103/PhysRevD.93.043013}, \href
  {http://adsabs.harvard.edu/abs/2016PhRvD..93d3013L} {93, 043013}

\bibitem[\protect\citeauthoryear{{Lupton}, {Gunn}  \& {Szalay}}{{Lupton}
  et~al.}{1999}]{luptitude}
{Lupton} R.~H.,  {Gunn} J.~E.,   {Szalay} A.~S.,  1999, \mn@doi [\aj]
  {10.1086/301004}, \href {http://adsabs.harvard.edu/abs/1999AJ....118.1406L}
  {118, 1406}

\bibitem[\protect\citeauthoryear{{Majumdar}, {Mellema}, {Datta}, {Jensen},
  {Choudhury}, {Bharadwaj}  \& {Friedrich}}{{Majumdar}
  et~al.}{2014}]{Suman2014}
{Majumdar} S.,  {Mellema} G.,  {Datta} K.~K.,  {Jensen} H.,  {Choudhury} T.~R.,
   {Bharadwaj} S.,   {Friedrich} M.~M.,  2014, \mn@doi [\mnras]
  {10.1093/mnras/stu1342}, \href
  {http://adsabs.harvard.edu/abs/2014MNRAS.443.2843M} {443, 2843}

\bibitem[\protect\citeauthoryear{{Majumdar}, {Pritchard}, {Mondal},
  {Watkinson}, {Bharadwaj}  \& {Mellema}}{{Majumdar} et~al.}{2018}]{suman2018}
{Majumdar} S.,  {Pritchard} J.~R.,  {Mondal} R.,  {Watkinson} C.~A.,
  {Bharadwaj} S.,   {Mellema} G.,  2018, \mn@doi [\mnras]
  {10.1093/mnras/sty535}, \href
  {http://adsabs.harvard.edu/abs/2018MNRAS.476.4007M} {476, 4007}

\bibitem[\protect\citeauthoryear{McKay}{McKay}{1979}]{LatinHypercube}
McKay M.~D. e.~a.,  1979, Technometrics, pp vol. 21, no. 2, pp. 239–245

\bibitem[\protect\citeauthoryear{{McQuinn}, {Lidz}, {Zahn}, {Dutta},
  {Hernquist}  \& {Zaldarriaga}}{{McQuinn} et~al.}{2007}]{McQuinn2007}
{McQuinn} M.,  {Lidz} A.,  {Zahn} O.,  {Dutta} S.,  {Hernquist} L.,
  {Zaldarriaga} M.,  2007, \mn@doi [\mnras] {10.1111/j.1365-2966.2007.11489.x},
  \href {http://adsabs.harvard.edu/abs/2007MNRAS.377.1043M} {377, 1043}

\bibitem[\protect\citeauthoryear{{Mellema}, {Iliev}, {Alvarez}  \&
  {Shapiro}}{{Mellema} et~al.}{2006}]{c2ray}
{Mellema} G.,  {Iliev} I.~T.,  {Alvarez} M.~A.,   {Shapiro} P.~R.,  2006,
  \mn@doi [\na] {10.1016/j.newast.2005.09.004}, \href
  {http://adsabs.harvard.edu/abs/2006NewA...11..374M} {11, 374}

\bibitem[\protect\citeauthoryear{{Mellema} et~al.,}{{Mellema}
  et~al.}{2013}]{ska}
{Mellema} G.,  et~al., 2013, \mn@doi [Experimental Astronomy]
  {10.1007/s10686-013-9334-5}, \href
  {http://adsabs.harvard.edu/abs/2013ExA....36..235M} {36, 235}

\bibitem[\protect\citeauthoryear{Mesinger \& Furlanetto}{Mesinger \&
  Furlanetto}{2007}]{Mesinger2007}
Mesinger A.,  Furlanetto S.,  2007, The Astrophysical Journal, 669, 663

\bibitem[\protect\citeauthoryear{{Mesinger}, {Furlanetto}  \& {Cen}}{{Mesinger}
  et~al.}{2011}]{21cmFast}
{Mesinger} A.,  {Furlanetto} S.,   {Cen} R.,  2011, \mn@doi [\mnras]
  {10.1111/j.1365-2966.2010.17731.x}, \href
  {http://adsabs.harvard.edu/abs/2011MNRAS.411..955M} {411, 955}

\bibitem[\protect\citeauthoryear{{Patil} et~al.,}{{Patil} et~al.}{2017}]{LOFAR}
{Patil} A.~H.,  et~al., 2017, \mn@doi [\apj] {10.3847/1538-4357/aa63e7}, \href
  {http://adsabs.harvard.edu/abs/2017ApJ...838...65P} {838, 65}

\bibitem[\protect\citeauthoryear{Pedregosa et~al.,}{Pedregosa
  et~al.}{2011}]{scikit-learn}
Pedregosa F.,  et~al., 2011, Journal of Machine Learning Research, 12, 2825

\bibitem[\protect\citeauthoryear{{Pober}, {Greig}  \& {Mesinger}}{{Pober}
  et~al.}{2016}]{Pober2016}
{Pober} J.~C.,  {Greig} B.,   {Mesinger} A.,  2016, \mn@doi [\mnras]
  {10.1093/mnrasl/slw156}, \href
  {http://adsabs.harvard.edu/abs/2016MNRAS.463L..56P} {463, L56}

\bibitem[\protect\citeauthoryear{{Press} \& {Schechter}}{{Press} \&
  {Schechter}}{1974}]{PressSchechter1974}
{Press} W.~H.,  {Schechter} P.,  1974, \mn@doi [\apj] {10.1086/152650}, \href
  {http://adsabs.harvard.edu/abs/1974ApJ...187..425P} {187, 425}

\bibitem[\protect\citeauthoryear{{Pritchard} \& {Loeb}}{{Pritchard} \&
  {Loeb}}{2012}]{PritchardLoeb2012}
{Pritchard} J.~R.,  {Loeb} A.,  2012, \mn@doi [Reports on Progress in Physics]
  {10.1088/0034-4885/75/8/086901}, \href
  {http://adsabs.harvard.edu/abs/2012RPPh...75h6901P} {75, 086901}

\bibitem[\protect\citeauthoryear{Rasmussen \& Williams}{Rasmussen \&
  Williams}{2006}]{Rasmussen}
Rasmussen Williams 2006, Gaussian Processes for Machine Learning.
The MIT Press

\bibitem[\protect\citeauthoryear{Rumelhart, Hinton  \& Williams}{Rumelhart
  et~al.}{1986}]{Rumelhart1986}
Rumelhart D.~E.,  Hinton G.~E.,   Williams R.~J.,  1986, Nature, 323, 533

\bibitem[\protect\citeauthoryear{{Santos}, {Ferramacho}, {Silva}, {Amblard}  \&
  {Cooray}}{{Santos} et~al.}{2010}]{Simfast21}
{Santos} M.~G.,  {Ferramacho} L.,  {Silva} M.~B.,  {Amblard} A.,   {Cooray} A.,
   2010, \mn@doi [\mnras] {10.1111/j.1365-2966.2010.16898.x}, \href
  {http://adsabs.harvard.edu/abs/2010MNRAS.406.2421S} {406, 2421}

\bibitem[\protect\citeauthoryear{{Schmit} \& {Pritchard}}{{Schmit} \&
  {Pritchard}}{2018}]{Schmit2018}
{Schmit} C.~J.,  {Pritchard} J.~R.,  2018, \mn@doi [\mnras]
  {10.1093/mnras/stx3292}, \href
  {http://adsabs.harvard.edu/abs/2018MNRAS.475.1213S} {475, 1213}

\bibitem[\protect\citeauthoryear{{Semelin}, {Eames}, {Bolgar}  \&
  {Caillat}}{{Semelin} et~al.}{2017}]{Semelin2017}
{Semelin} B.,  {Eames} E.,  {Bolgar} F.,   {Caillat} M.,  2017, \mn@doi
  [\mnras] {10.1093/mnras/stx2274}, \href
  {http://adsabs.harvard.edu/abs/2017MNRAS.472.4508S} {472, 4508}

\bibitem[\protect\citeauthoryear{{Sheth}, {Mo}  \& {Tormen}}{{Sheth}
  et~al.}{2001}]{Sheth2001}
{Sheth} R.~K.,  {Mo} H.~J.,   {Tormen} G.,  2001, \mn@doi [\mnras]
  {10.1046/j.1365-8711.2001.04006.x}, \href
  {http://adsabs.harvard.edu/abs/2001MNRAS.323....1S} {323, 1}

\bibitem[\protect\citeauthoryear{{Shimabukuro} \& {Semelin}}{{Shimabukuro} \&
  {Semelin}}{2017}]{Shimabukuro2017}
{Shimabukuro} H.,  {Semelin} B.,  2017, \mn@doi [\mnras]
  {10.1093/mnras/stx734}, \href
  {http://adsabs.harvard.edu/abs/2017MNRAS.468.3869S} {468, 3869}

\bibitem[\protect\citeauthoryear{{Shimabukuro}, {Yoshiura}, {Takahashi},
  {Yokoyama}  \& {Ichiki}}{{Shimabukuro}
  et~al.}{2016}]{Shimabukuro2016Bispectrum}
{Shimabukuro} H.,  {Yoshiura} S.,  {Takahashi} K.,  {Yokoyama} S.,   {Ichiki}
  K.,  2016, \mn@doi [\mnras] {10.1093/mnras/stw482}, \href
  {http://adsabs.harvard.edu/abs/2016MNRAS.458.3003S} {458, 3003}

\bibitem[\protect\citeauthoryear{{Sobacchi} \& {Mesinger}}{{Sobacchi} \&
  {Mesinger}}{2014}]{Sobacchi&Mesinger2014}
{Sobacchi} E.,  {Mesinger} A.,  2014, \mn@doi [\mnras] {10.1093/mnras/stu377},
  \href {http://adsabs.harvard.edu/abs/2014MNRAS.440.1662S} {440, 1662}

\bibitem[\protect\citeauthoryear{{Tingay} et~al.,}{{Tingay} et~al.}{2013}]{mwa}
{Tingay} S.~J.,  et~al., 2013, \mn@doi [\pasa] {10.1017/pasa.2012.007}, \href
  {http://adsabs.harvard.edu/abs/2013PASA...30....7T} {30, e007}

\bibitem[\protect\citeauthoryear{Titsias}{Titsias}{2009}]{Titsias09}
Titsias M.,  2009, in Proceedings of the Twelth International Conference on
  Artificial Intelligence and Statistics. PMLR, Hilton Clearwater Beach Resort,
  Clearwater Beach, Florida USA, pp 567--574, \url
  {http://proceedings.mlr.press/v5/titsias09a.html}

\bibitem[\protect\citeauthoryear{{Watkinson}, {Majumdar}, {Pritchard}  \&
  {Mondal}}{{Watkinson} et~al.}{2017}]{Watkinson2017}
{Watkinson} C.~A.,  {Majumdar} S.,  {Pritchard} J.~R.,   {Mondal} R.,  2017,
  \mn@doi [\mnras] {10.1093/mnras/stx2130}, \href
  {http://adsabs.harvard.edu/abs/2017MNRAS.472.2436W} {472, 2436}

\bibitem[\protect\citeauthoryear{{Watkinson}, {Giri}, {Ross}, {Dixon}, {Iliev},
  {Mellama}  \& {Pritchard}}{{Watkinson} et~al.}{2018}]{Watkinson2018}
{Watkinson} C.~A.,  {Giri} S.~K.,  {Ross} H.~E.,  {Dixon} K.~L.,  {Iliev}
  I.~T.,  {Mellama} G.,   {Pritchard} J.~R.,  2018, preprint, \href
  {http://adsabs.harvard.edu/abs/2018arXiv180802372W} {} (\mn@eprint {arXiv}
  {1808.02372})

\bibitem[\protect\citeauthoryear{Werbos}{Werbos}{1982}]{backprop}
Werbos P.~J.,  1982, in Drenick R.~F.,  Kozin F.,  eds, System Modeling and
  Optimization. Springer Berlin Heidelberg, Berlin, Heidelberg, pp 762--770

\bibitem[\protect\citeauthoryear{Zahn, Lidz, McQuinn, Dutta, Hernquist,
  Zaldarriaga  \& Furlanetto}{Zahn et~al.}{2007}]{Zahn2007}
Zahn O.,  Lidz A.,  McQuinn M.,  Dutta S.,  Hernquist L.,  Zaldarriaga M.,
  Furlanetto S.~R.,  2007, The Astrophysical Journal, 654, 12

\bibitem[\protect\citeauthoryear{{Zel'dovich}}{{Zel'dovich}}{1970}]{Zeldovich1970}
{Zel'dovich} Y.~B.,  1970, \aap, \href
  {http://adsabs.harvard.edu/abs/1970A%26A.....5...84Z} {5, 84}

\makeatother
\end{thebibliography}
\bibliographystyle{mnras}

\begin{appendices}

\section{Simulation parameters}
\label{appendix:parameters}
We list all relevant user-changeable parameters used for all \cmfast{} and \simfast{} simulations in this paper. For further descriptions of these parameters see \mcite{21cmFast} and \mcite{Simfast21}. We exclude parameters relating to spin temperature calculations since we did no use this functionality.

\needspace{22\baselineskip}
\subsection{Cosmology}
\begin{tabular}{c c}
 \hline
 \textbf{Parameter} & \textbf{Value} \\
 \hline
 $\sigma_8$ & 0.810 \\
 Hubble $h$ & 0.710 \\
 $\Omega_{\mathrm{M}}$ & 0.270 \\
 $\Omega_\Lambda$ & 0.730 \\
 $\Omega_{\mathrm{b}}$ & 0.046 \\
 $\Omega_{\mathrm{n}}$ & 0.0 \\
 $\Omega_{\mathrm{k}}$ & 0.0 \\
 $\Omega_{\mathrm{R}}$ & 0.0 \\
 $\Omega_{\mathrm{tot}}$ & 1.0 \\
 $Y_{\mathrm{He}}$ & 0.245 \\
 $n_{\mathrm{s}}$ & 0.960 \\
 Sheth-Tormen $b$ & 0.34 \\
 Sheth-Tormen $c$ & 0.81 \\
 Helium II $z_{\mathrm{reion}}$ & 3 \\
 Maximum Redshift  & 17.00 \\
 Minimum Redshift  & 8.00 \\
 Redshift Step & 1.50 \\
 Simulation Length & 500.00 \\
 Star Formation Rate & 0.025 \\
 Velocity Component & 3 \\ 
 Critical Overdensity & 1.680 \\
 \hline
\end{tabular}

\needspace{8\baselineskip}
\subsection{\simfast{} reionization parameters}
\begin{supertabular}{ c c } 
 \hline
 \textbf{Parameter Name} & \textbf{Value} \\
 \hline
  use\_camb\_matterpower & False \\
  use\_fcoll & True \\
  halo\_Rmax & 40 \\
  halo\_Mmin & Various \\
  Ion\_eff & Various \\
  bubble\_Rmax & Various \\
  use\_Lya\_xrays & False \\
 \hline
\end{supertabular}

\needspace{6\baselineskip}
\subsection{\cmfast{} reionization parameters}
\begin{supertabular}{ c c } 
 \hline
 \textbf{Parameter Name} & \textbf{Value} \\
 \hline
   ION\_M\_MIN & Various \\
  ION\_Tvir\_MIN & -1 (off) \\
  HII\_EFF\_FACTOR & Various \\
  EFF\_FACTOR\_PL\_INDEX & 0 \\
  R\_BUBBLE\_MAX & Various \\
  \hline
\end{supertabular}

\subsection{Other \cmfast{} parameters}
\begin{supertabular}{ c c } 
 \hline
 \textbf{Parameter Name} & \textbf{Value} \\
 \hline
  P\_CUTOFF & 0 \\ 
  M\_WDM & 2 \\ 
  g\_x & 1.5 \\  
  INHOMO\_RECO & 0 \\
  ALPHA\_UVB & 5 \\
  t\_STAR & 0.5 \\
  EVOLVE\_DENSITY\_LINEARLY & 0 \\
  SMOOTH\_EVOLVED\_DENSITY\_FIELD & 1 \\
  R\_smooth\_density & 0.2 \\
  SECOND\_ORDER\_LPT\_CORRECTIONS & 0 \\
  HII\_ROUND\_ERR & 1e-3 \\
  FIND\_BUBBLE\_ALGORITHM & 1 \\
  R\_BUBBLE\_MIN & L\_FACTOR*1 \\
  USE\_HALO\_FIELD & 0 \\
  N\_POISSON & -1 \\
  T\_USE\_VELOCITIES & 1 \\
  MAX\_DVDR & 0.2 \\
  DIMENSIONAL\_T\_POWER\_SPEC & 0 \\
  DELTA\_R\_FACTOR & 1.1 \\
  DELTA\_R\_HII\_FACTOR & 1.1 \\
  R\_OVERLAP\_FACTOR & 1.0 \\
  DELTA\_CRIT\_MODE & 1 \\
  HALO\_FILTER & 0 \\
  HII\_FILTER & 1 \\
  OPTIMIZE & 0 \\
  OPTIMIZE\_MIN\_MASS & 1e11 \\
  SIZE\_RANDOM\_SEED & -23456789 \\
  LOS\_RANDOM\_SEED & -123456789 \\
  USE\_TS\_IN\_21CM & 0 \\
  CLUMPING\_FACTOR & >50 \\
  Pop & 2 \\
  Pop2\_ion & 4361 \\
 \hline
\end{supertabular}

\end{appendices}

\bsp	
\label{lastpage}
\end{document}